\documentclass[a4paper]{article}

\usepackage{graphicx}
\usepackage{amssymb,amsmath}
\usepackage{xcolor}
\usepackage{subfigure}
\usepackage{amsmath}
\usepackage{booktabs}
\usepackage{tabularx}
\usepackage{placeins}
\usepackage{url}
\usepackage{authblk}
\usepackage[title]{appendix}

\begin{document}
\title{Classifying regions of high model error within a data-driven RANS closure: Application to wind turbine wakes}
\author[1]{Julia Steiner (corresponding author)}
\author[1]{Richard P. Dwight}
\author[1]{Axelle Vir\'{e}}
\affil[1]{Aerodynamics \& Wind energy, Aerospace faculty, Kluyverweg 1, 2629 HS Delft, Netherlands}
\affil[ ]{\textit {\{j.steiner,r.p.dwight,a.c.vire\}@tudelft.nl}}

\maketitle

\begin{abstract}
  Data-driven Reynolds-Averaged Navier-Stokes (RANS) turbulence closures are increasing seen as a viable alternative to general-purpose RANS closures, when LES reference data is available -- also in wind-energy.  Parsimonious closures with few, simple terms have advantages in terms of stability, interpret-ability, and execution speed.  However experience suggests that closure model corrections need be made only in limited regions -- e.g.\ in the near-wake of wind turbines and not in the majority of the flow.  A parsimonious model therefore must find a middle ground between precise corrections in the wake, and zero corrections elsewhere.  We attempt to resolve this impasse by introducing a classifier to identify regions needing correction, and only fit and apply our model correction there.  We observe that such a classifier (which must be computed only from RANS-available quantities) is straightforward to construct, and accurate in operation.  We further observe that the correction models are significantly simpler (with fewer terms) when limited to the identified regions than otherwise, but have similar or better accuracy.  We apply our framework to three flows consisting of multiple wind-turbines in neutral conditions with interacting wakes.
\end{abstract}

\section{Introduction}\label{sec:intro}
Aerodynamic models for wind farms are essential to optimize energy yield and turbine loading both during the design and the operational stage of wind farms.  The simplest models are algebraic engineering models, and the most complex Large-Eddy Simulations (LES) \cite{stevens2017}.  While the former do not give meaningful results if strong wake interaction is present, the latter are too expensive to be used for engineering purposes \cite{breton2017,ghaisas2017}.  Reynolds-Averaged Navier-Stokes (RANS) offers a possible middle way: they require about two orders of magnitude less computational time than LES, but have significant accuracy deficits caused by the modelling of turbulence closure.

For quasi-steady atmospheric conditions around wind farms, the most commonly used RANS model is the $k-\varepsilon$ model. However, this model has crippling structural shortcomings. It over-predicts the eddy viscosity in the near wake which leads to an accelerated wake recovery, and it fails to account for the effects of turbulence anisotropy \cite{sanderse2011}, and the direct effect of the turbine on the turbulence mean quantities is not modeled \cite{rethore2009}. Many modifications have been proposed in literature, but the improvements are test-case specific, require non-general tuning parameters, are not numerically robust, and atmospheric stratification is not yet considered \cite{gomez2005,elkasmi2008,rethore2009,prospathopoulos2011,cabezon2011,laan2015,laan2018}. Additionally, most of these models do not directly consider the effect of actuator forcing on the turbulence equations.  The most successful modification proposed so far is the  $k-\varepsilon-f_P$ model from van den Laan et al.~\cite{laan2018}, who uses an eddy-viscosity limiter that is only active in regions with high-velocity gradients.  A summary of possible modifications can be found in our previous publication \cite{steiner2020b}.  The perspective for further improvements by expert modellers is in our opinion limited, given the complexity of the modelling task -- especially when stratification is considered.

It is for this reason that we propose data-driven modelling as an tool for devising more accurate closures.  In our previous work~\cite{steiner2020a,steiner2020b} we extended the data-driven framework Sparse Regression of Turbulent Stress Anisotropy (SpaRTA) first introduced by Schmelzer et al.~\cite{schmelzer2019}.  The framework introduces two nonlinear corrections to the baseline $k-\varepsilon$ or $k-\omega$ equations: (i) an anisotropy correction and (ii) an additive correction to the transport equation for the turbulent kinetic energy. This has the benefit of correcting both the directionality and the magnitude of the Reynolds stress tensor (RST), as well as accounting for model-form errors in the transport equation for $k$. Furthermore SpaRTA uses deterministic symbolic regression, for which the search space is constrained towards parsimonious algebraic models using sparsity-promoting regression techniques \cite{brunton2016,rudy2017}.

While the framework worked well for our turbine wake-interaction problems~\cite{steiner2020a}, leading to significantly improved wake-evolution predictions, the experience showed that the closure model corrections were necessary only in very limited spatial regions of the broader flow.  In particular, while the majority of the domain consisted of an undisturbed Atmospheric Boundary Layer (ABL), corrections to the baseline $k-\varepsilon$ model were needed only in limited regions -- specifically the wake, and especially the near-wake.  Our parsimonious SpaRTA model therefore was forced to find a middle ground between precise corrections in the wake, and zero corrections elsewhere.  it achieved this will terms that cancelled in many circumstances.

To resolve this issue -- and the novelty of this work -- is the addition of a logistic classifier to the SpaRTA framework. A classifier is a function that yields values between zero and one, and is used here to switch the closure corrections off and on, so they are active only where needed, and elsewhere the standard $k-\varepsilon$ model is used.  This is analogous to sensors in traditional closures, which detect specific physical effects and active relevant terms only locally.  Our classifier is trained based on the magnitude of the required model correction (a quantity derived from the LES data), rather than on metrics estimating the significance of RANS modelling assumptions as in~\cite{ling2015,gorle2015}.  As such the classifier directly pertains to the need-for-correction.  The logistic classifier by its nature gives a smooth transition between "off" and "on", reducing spurious effects due to switching, and as a side-effect, both training and prediction computational costs are reduced (as a result of reduced data and simpler models respectively).

While classifiers have been explored in the context of RANS modelling before, e.g.~\cite{ling2015}, this work is -- to our knowledge -- the first example of a learnt classifier forming an integral part of a RANS closure.  The classifier is itself parsimonious (being based on symbolic regression), straightforward to construct, and is shown to generalize well.  The correction models obtained in combination with the classifier are significantly simpler (with fewer terms) than correction models based on the full field, but have similar or better accuracy in a predictive setting.

While there are a multitude of publications on data-driven turbulence modeling with various approaches as summarized in Duraisamy et al.~\cite{duraisamy2019}, only few use classifiers or markers to identify regions in the flow field with high uncertainty due to the turbulence model.  Gorl\'{e} et al. \cite{gorle2015} developed a simple nonlinear marker for RANS simulations to identify regions in which the flow field deviates from parallel shear flow. The results showed good agreement between a positive marker and an inaccurate prediction of the Reynolds stress divergence for the two test cases that they used.  Ling et al. \cite{ling2015} defined three separate markers that pertain to different ways in which the Boussinesq hypothesis fails: (i) the negativity of the eddy viscosity, (ii) turbulence anisotropy, and (iii) the difference between a linear and nonlinear eddy-viscosity model prediction.  The markers were derived by solving a classification problem using different supervised machine-learning approaches, namely Support Vector Machines (SVMs), Adaboost decision trees, and Random Forests (RFs).

However, none of these publications integrate these markers with either a turbulence correction or more accurate turbulence models in regions with positive indication. The authors are aware of one publication publication by Longo et al. \cite{longo2017} where the marker from Gorl\'{e} et al.\ is used to blend a LEVM model with a NLEVM in regions of non-parallel shear flow around buildings.  A blending function is used to further smooth the marker properties, because the marker itself can have very sharp gradients. This approach has some similarity with the one presented in this paper: we also uses a nonlinear eddy-viscosity model (NLEVM) in combination with a marker that makes sure that the nonlinear corrections are only applied selectively.  However both the marker and correction are fully hand-designed, whereas ours result from a data-driven approach targeted at a specific class of flows.

The general approach of selectively modifying the closure model depending on the local flow properties has parallels in the Generalized $k-\omega$ (GEKO) models of Mentor~\cite{menter2020}.  In wind-energy the $k-\varepsilon-f_P$ model uses a kind of classifier which scales the eddy viscosity.  Both these are hand-designed models.

This publication is structured as follows.  In Section \ref{sec:methodology} we specify the methodology. Additive model-form error terms within the $k-\varepsilon$ LEVM model are identified via the introduction of corrections to the stress-strain relation and the turbulence transport equations. The $k$-corrective-frozen-RANS approach to identify the optimal model correction is explained, and the target for the classifier is defined.  The modelling of both the correction term and the classifier using an elastic net is introduced.  In Section \ref{sec:results}, the results of the frozen approach, the training, and cross-validation of the classifier and the correction terms, as well as the inclusion of the weighted correction terms in the flow solver are displayed. Some thoughts on numerical stability are also presented.  A comparison between models derived with and without the classifier is also shown. Finally, conclusions are drawn in Section \ref{sec:conclusions}.

\section{Methodology}\label{sec:methodology}
The ordering of this section follows the sequence of our turbulence modeling chain. Firstly (Section~\ref{s:les}) the ground-truth database is generated using an LES solver. Secondly (Section~\ref{s:frozen}) the frozen approach is introduced to derive optimal corrective fields to the RANS equations.  Thirdly (Section~\ref{s:classy}) the target of the classifier is introduced.  Finally, in Section~\ref{s:sparsereg} a sparse symbolic regression procedure is described with which generalized algebraic expressions of the optimal correction fields and the classifier can be constructed. All these steps result in a new turbulence closure model which is capable of generalizing beyond the training data-set to similar test-cases.

\subsection{Ground-truth data generation (LES)}
\label{s:les}
\begin{figure}[h!]
\centering
\includegraphics[clip,trim={0.2\textwidth} {0.1\textwidth} {0.2\textwidth} {0.1\textwidth},width=0.4\textwidth]{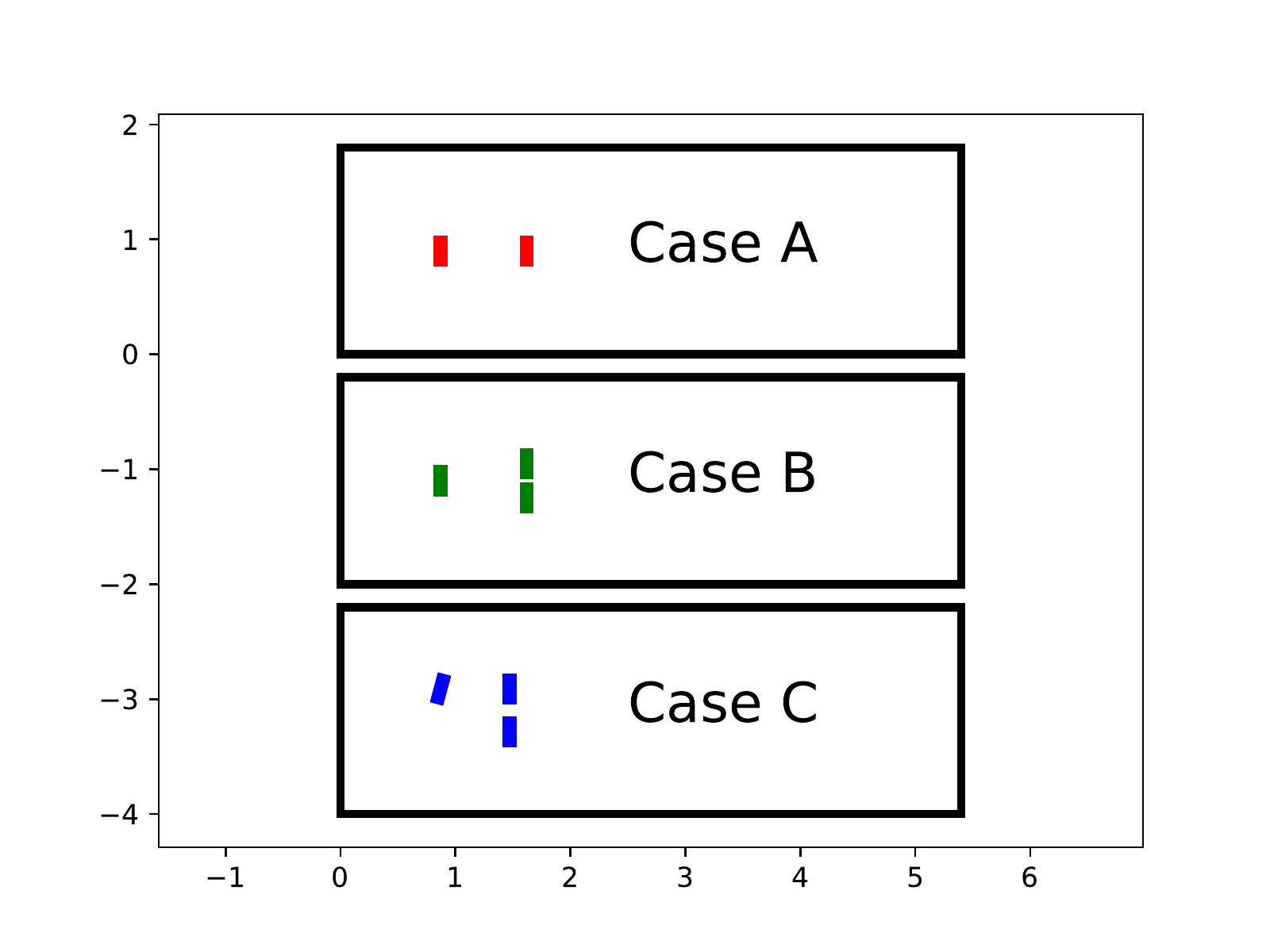}
\caption{Case constellation, turbine diameter is to scale.}\label{fig:cases}
\end{figure}

The foundation of data-driven modeling is a good database.  Our database comprised three different cases referred to as cases A, B \& C. Figure~\ref{fig:cases} shows the three cases considered in this paper. All cases use the same surface roughness and inflow velocity profile, but varying turbine constellations. The turbine and inflow properties were taken from the wind-tunnel experiment of Chamorro and Port\'{e}-Agel \cite{chamorro2010}. Table~\ref{tab:cases} presents the parameters related to the inflow profiles, turbine dimension and domain size.

\begin{table}[]
\centering
\begin{minipage}{18pc}
\centering
\begin{tabular}{@{}*{7}{l}}
\hline
\textbf{Turbine} & \\
\hline
Diameter&$D=0.15\text{m}$\\
Hub height&$h_{hub}=0.125\text{m}$\\
Rotation speed &$\Omega=1190\text{rpm}$\\
\hline
\textbf{Inflow boundary} & \textbf{layer} \\
\hline
Velocity &$U\left(h_{hub}\right)=2.2\text{m/s}$\\
Turbulence intensity &$\sigma_U\left(h_{hub}\right) =1.0 \%$\\
\hline
\textbf{Mesh} & \\
\hline
Domain size &  $5.4 \times 1.8 \times 0.46 \text{m}^3$ \\
Resolution & $360 \times 120 \times 64 $ \\
\hline
\end{tabular}\caption{Case setup paramters}\label{tab:cases}
\end{minipage}
\begin{minipage}{18pc}
\centering
\begin{tabular}{@{}*{7}{l}}
\hline
\textbf{WALE model} & \\
\hline
$C_e$ & $0.93$\\
$C_k$ & $0.0673$\\
$C_w$ & $0.325$\\
\hline
\textbf{$k-\varepsilon$ model} &  \\
\hline
$C_\mu$ & $0.03$\\
$C_{\varepsilon 1}$ & $1.42$\\
$C_{\varepsilon 2}$ & $1.92$\\
$\sigma_{\varepsilon}$ & $1.3$\\
$\sigma_{k}$ & $1.3$\\
\hline
\end{tabular}\caption{Turbulence model parameters}\label{tab:turbParameters}
\end{minipage}
\end{table}

For the CFD model, \texttt{OpenFOAM-6.0} was used in conjunction with the \texttt{SOWFA-6} toolbox \cite{sowfa}. For the RANS solver, a modified $k-\varepsilon$ model is the baseline closure; for the LES solver, the WALE model was used to model the unresolved scales \cite{nicoud1999,sanz2017}. Table~\ref{tab:turbParameters} presents the specific closure coefficients used here. Actuator-disc models of the wind turbines are used.  In the LES we could potentially have used actuator-line models, but for consistency with RANS we prefer disc models.  Details of the numerical setup such as boundary conditions and mesh resolution, as well as a validation of the CFD models with respect to wind tunnel measurements, can all be found in \cite{steiner2020b}.

\FloatBarrier

\subsection{Optimal RANS corrections using the frozen approach}
\label{s:frozen}
The basic premise of this section is as follows: take an LES time-averaged flow-field of a statistically stationary flow, including mean velocity $U^\star$, turbulent kinetic energy $k^\star$, and Reynolds stresses $\tau_{ij}^\star$ (where an LES quantity is denoted by a $\star$).  If we inject these quantities into the $k-\varepsilon$ equations with appropriate B.C.s, the only remaining unknown is the turbulence dissipation rate $\varepsilon$.  We can solve the $\varepsilon$-equation (from $k-\varepsilon$) to obtain an approximation to the dissipation rate, but the other equations (for $k$ and $U$) will not be satisfied.  This is due to the modelling assumptions required to obtain these equations - notably the Boussinesq assumption, and the modelling of the various terms in the $k$ equation.  As a direct consequence: solving the $k-\varepsilon$ equations can not lead us to the LES solution.

If we wish to have a solution that corresponds to the LES mean flow, it is therefore necessary to modify the equations such that when LES data is injected they {\it are} satisfied.  The way we do this is by adding spatially varying {\it corrective fields}, and since both the momentum equation and the $k$-equation are not satisfied, we need to add corrections to both these equations.  This can be done in a number of ways, we choose an additive correction $\textcolor{blue}{\check{R}(\mathbf{x})}$ to the $k$-equation:

\begin{equation}\label{eq:kEpsilonABLk}
        \frac{D k^\star}{D t} = \mathcal{P}_k^\star + \textcolor{blue}{\check{R}(\mathbf{x})}
         - \varepsilon +\frac{\partial }{\partial x_j} \left[  \left(\nu +\nu_t/\sigma_k \right) \frac{\partial {k^\star}}{\partial x_j} \right],
\end{equation}

and interpreting this as a modification to the production leads to a corresponding change in the $\varepsilon$ equation:

\begin{equation}\label{eq:kEpsilonABLepsilon}
        \frac{D \varepsilon}{D t} = \left[C_{\varepsilon 1} \left(\mathcal{P}_k^\star+
                                        \textcolor{blue}{\check{R}(\mathbf{x})} \right)
                                         -C_{\varepsilon 2} \varepsilon  \right] \cdot                                             \frac{\varepsilon}{k^\star}
                                         +\frac{\partial }{\partial x_j} \left[  \left(\nu + \nu_t/\sigma_\varepsilon\right) \frac{\partial \varepsilon}{\partial x_j} \right],
\end{equation}

where the production term is known from the LES data

\begin{equation}
\mathcal{P}_k^\star := 2 k^\star b_{ij}^\star \frac{\partial U_i^\star}{\partial x_j}.
\end{equation}
The above coupled equations consist of a PDE for $\varepsilon$ and an algebraic expression for $\textcolor{blue}{\check{R}}$, and may be solved iteratively yielding the unknown fields $\varepsilon$ and $\textcolor{blue}{\check{R}}$.

We must also address the momentum equation, which we do by a correction to the Boussinesq approximation:
\begin{equation}
\label{eq:bij}
b_{ij}^\star := \frac{\tau_{ij}^\star}{2 k^\star} - \frac{1}{3} \delta_{ij} = -\frac{\nu_t}{k^\star} S_{ij}^\star + \textcolor{red}{\check{b}_{ij}^\Delta(\mathbf{x})}.
\end{equation}
so that the tensor-field $\textcolor{red}{\check{b}_{ij}^\Delta(\mathbf{x})}$ can be computed algebraically from \eqref{eq:bij} using the definition of the eddy-viscosity $\nu_t := C_\mu k^{\star 2}/\varepsilon$.  The resulting fields satisfy the modified $k-\varepsilon$ equations, with the LES data as an exact solution.

\subsection{Specification of the classifier}
\label{s:classy}
The corrective fields $\textcolor{red}{\check{b}_{ij}^\Delta(\mathbf{x})}$ and $\textcolor{blue}{\check{R}(\mathbf{x})}$ defined above are non-zero everywhere in the field, but generally small outside of the wake.  As such we define a marker given which correction is likely required as:

\begin{equation}
\label{eq:clfOptimal}
\sigma_{wake} =
\begin{cases}
    1 &
    \text{if }
    \left(
    \left\lvert \mathcal{P}_k^\Delta \right\rvert > 0.02
    \right)
    \cup
    \left(
    \left\lvert \frac{U-U_\mathrm{ABL}}{U_\mathrm{ABL}}\right\rvert > 0.05 \right) \\
   0 & \text{otherwise},
\end{cases}
\end{equation}

where $U_\mathrm{ABL}$ is the undisturbed boundary-layer velocity at a given height, and
\[
\mathcal{P}_k^\Delta := b_{ij}^\Delta \frac{\partial U_i}{\partial x_j}
\]
is the extra T.K.E.\ production due to $b_{ij}^\Delta$.  This marker is thus a combination of a indicator of significant anisotropy correction (with an effect on production), and a wake sensor.  The latter is specific to our application, and was found to be necessary, otherwise the area with a positive classifier was non-smooth. The former is applicable in a general setting.

Note that the evolution of the incoming ABL does not match exactly between RANS and LES, and hence an additional correction is required (a function of wall-distnace only), the details of which are not relevant here, see \cite{steiner2020b} for more information.

\subsection{Learning of the correction terms and the classifier}
\label{s:sparsereg}
The objective of this section is to take the corrective fields $\textcolor{red}{\check{b}_{ij}^\Delta(\mathbf{x})}$ and $\textcolor{blue}{\check{R}(\mathbf{x})}$, and the classifier $\sigma_{wake}(\mathbf{x})$, which are currently all functions of space for a specific test-case, and make (potentially) generalizable models for them in terms of the flow quantities available to RANS.  This is the point at which the methods of supervised machine learning are valuable.

The input features we use are as comprehensive as we can achieve - later sparse regression will eliminate features that are not informative.  We use an integrity basis based on the set $\{\textbf{S},{\boldsymbol \Omega},\textbf{A}_p,\textbf{A}_k\}$ where $\textbf{S}:=\frac{1}{2}(\nabla U + \nabla U^T)$,
${\boldsymbol \Omega} := \frac{1}{2}(\nabla U - \nabla U^T)$,
$\textbf{A}_p = -I \times \nabla p$ and $\textbf{A}_k = -I \times \nabla k$.  This is original Pope basis~\cite{pope1975} augmented with pressure- and $k$-gradients.  We used all 47 invariants $\mathbf{I} := [I_1,\dots,I_{47}]$ of this basis as features.  In addition we supplement the feature set with 11 non-dimensionalized physical features such as actuator forcing, $\mathbf{q} := [q_1,\dots,q_{11}]$.   We use all features when approximating both corrective fields and the classifier.  The full list of features can be found in tables \ref{tab:physicalFeaturesAll} and \ref{tab:invariantsAll} in Appendix~\ref{app:Features}.

Where approximating $b^\Delta$, we employ the basis tensors $T_{ij}^{(n)}$, thereby guaranteeing Galilean and rotational invariance:
\[
b^\Delta_{ij} \simeq \sum_{n=1}^{10} T_{ij}^{(n)} \alpha_n(\mathbf{I},\mathbf{q}).
\]
An analogous modelling approach is taken for $\textcolor{blue}{R}$:

\begin{equation}
\textcolor{blue}{R} \simeq 2 k \frac{\partial u_i}{\partial x_j}
\sum_{n=1}^{10} T_{ij}^{(n)} \beta_n \left( \mathbf{I},\mathbf{q}\right)
+ \varepsilon \cdot \gamma \left(\mathbf{I},\mathbf{q} \right),
\label{eq:R}
\end{equation}

where we allow two types of terms are used: one that mirrors a correction to the turbulence production, and one that represents a more general scalar correction. The scalar term is scaled with the turbulent dissipation rate $\varepsilon$, in order to keep the term dimensionally correct.  The motivation for these two separate corrections is to capture both errors in the production term itself (which in many flows is the dominant term), as well as other model-form errors, notably the omission of the effect of the rotor forcing on the turbulence.

For the classifier the sigmoid function $\sigma$ is used in conjunction with a scalar function $\delta$ that is modelled in the same way as the scalar function for the correction terms:
\begin{equation}
\label{eq:sigma}
    \textcolor{black}{\sigma_{wake}} \left(\mathbf{I},\mathbf{q}\right)  =
    \frac{1}
    {1+\exp^{-\delta \left( \mathbf{I},\mathbf{q} \right)}} = \sigma \left[ \delta\left(\mathbf{I},\mathbf{q}\right)\right]
    \ \ \text{with} \ \ \sigma^{wake} \in \left\{0,1\right\} .
\end{equation}
The sigmoid function $\sigma$ forces the numeric range of the values to be between $0$ and $1$, even if the range of the scalar function  $\delta$ is different from that in the training data-set.

The formulation for the scalar functions  $\alpha_n$, $\beta_n$, $\gamma$, and $\delta$  is based on the input feature set.  The $47 + 11 = 58$ input features are used to build a large library of $L$ candidate (basis) functions $(\ell_1, \dots \ell_L)$. This is done by recombining features with each other (up to a maximum of three features), and applying exponentiation by $\frac{1}{2}$ and $2$.  This already results in a library signficantly larger than the feature set.  Each scalar function is then represented as:
\begin{equation}
  \alpha_n(\textbf{I}, \textbf{q}) \simeq \sum_{k=0}^L \theta^n_k \ell_k(\textbf{I},\textbf{q}),
  \label{eq:alpha}
\end{equation}
i.e.\ a linear regression problem with coefficients $\theta$.  An elastic net is then used to identify an optimal regressor with sparsity (most of the coefficients are zero) \cite{Zou_2005}.   Logistic regression is convenient fpr the classfier due to the exponential character of the sigmoid function. The discrete counterpart of equation \eqref{eq:sigma} is
\[
\sigma_{wake} = \sigma \left[ C \cdot \theta^\sigma\right].
\]
where $C$ is the matrix consisting of all basis functions evaluated at all data points (corresponding to mesh-points of the LES training simulation).

The outline of the full procedure is detailed below - we focus attention on the classfier - the other terms are analogous:

\begin{enumerate}
\item{\textbf{Data reduction}}: Use the reference classifier $\sigma_{wake} \in \{0,1\}$ as a condition for inclusion in the training dataset.  Note: This step is not applicable for training the classifier, only training the correction terms given a classifier.
\item{\textbf{Preprocessing}}: Use a mutual-information criterion to reduce the feature set, then building the library; and reduce it by {\it cliqueing} (identifying and removing sets of multi-colinear functions).
\item{\textbf{Model discovery}}: Use an elastic net to identify important library functions.  By varying regularization parameters $\lambda$ and $\rho$, the result is an array of models with a variety of complexity and accuracy.  The optimization problem for the classifier is:
\begin{equation}
\min_{\theta^{\sigma}} \left[
  \sum_i \ln \left( \sigma \left[
  \sigma^{wake}_i C_i \Theta^\sigma \right]  \right)
  + \lambda \rho \left\| \theta^{\sigma} \right\|_1
  + 0.5 \lambda \left(1-\rho\right) \left\| \theta^{\sigma} \right\|_2^2
\right]
\end{equation}

\item \textbf{Remove unnecessary functions} from the library by eliminating all basis functions for which the corresponding $\theta=0$ for each of the models found in (ii).  The matrix $C \rightarrow \tilde{C}$ and $\Theta \rightarrow \tilde{\Theta}$ are also reduced.

\item{\textbf{Model calibration}} using Ridge regression to identify the magnitude of the model coefficients for the previously derived array of corrections.  Again a regularization parameter $\lambda_R$ is used to encourage small coefficients:

\begin{equation}
\min_{\theta^{\sigma}} \left[
  \sum_i \ln \left( \sigma \left[
  \sigma^{wake}_i C^\sigma_i \Theta^\sigma \right]  \right)
  + \lambda_r \left\| \theta^{\sigma} \right\|_2^2
\right]
\end{equation}
\end{enumerate}

The preprocessing step makes use of two probabilistic procedures: Mutual information (MI) \cite{Moon_1995,miSource} and cliqueing \cite{cliqueSource}. MI can identify nonlinear relations between input features and correction terms and can hence help reduce the input feature set. Cliqueing checks if there is multi-collinearity in the input library and is thus useful for discarding co-linear input functions. Both of these procedures are vital for bringing for making the learning procedure manageable for our dataset.

\section{Results and discussion}\label{sec:results}
This section shows the application of the proposed methodology to the previously described dataset. Resulting flow fields with the optimal and the learned isolated correction terms are shown in Sections \ref{sec:optimal} and \ref{sec:learning}, respectively. This is followed by a robustness analysis in Section \ref{sec:robustness}. Then, different combinations of correction terms and classifiers are compared in Section \ref{sec:learned}. Finally, a comparison is made between the models obtained with and without classifier in Section \ref{sec:wwoCL}.

In the following we consider RANS simulations with three different kinds of correction applied:

\begin{itemize}
    \item \textit{Frozen (or optimal)} refers to correction terms obtained from the frozen procedure of Section~\ref{s:frozen}.  In case of the classifier where a "frozen" term is not available, we refer to the training classifier identified by \eqref{eq:clfOptimal} as optimal.  A tilde is used to denote frozen terms: $\tilde{b}_{ij}^\Delta$, $\tilde{R}$ and $\tilde{\sigma}$.

     \item \textit{Fixed} refers to the correction term or classifier that results from applying a trained model to the LES flow field.  This is generally a good representation of the optimal correction, but includes errors due to the inability of the elastic-net to represent the optimal correction with the given features.  A hat is used to denote "fixed" terms:
     $\hat{b}_{ij}^\Delta$, $\hat{R}$ and $\hat{\sigma}$.

    \item \textit{Coupled} refers to a correction term that is a function of the flow field.  I.e.\ it changes as the flow-field changes, e.g.\ at every iteration of the flow solver.  In this sense it is a genuine turbulence model, operating independently of LES data.  The notation used for coupled terms is just $b_{ij}^\Delta$, $R$ and $\sigma$.
\end{itemize}

Various combinations of frozen, fixed and coupled correction terms are possible. For example, the correction $(\tilde{b}_{ij}^\Delta, \tilde{R}, \sigma)$ could be used to test how well a specific classifier operates independently of the performance of the anisotropy and production corrections.  Different model formulations are initially investigated like this to avoid coupling effects between different formulations. This is referred to in the following as "partially coupled".  Once a set of well working models is selected, all terms can be coupled simultaneously -- i.e.\ $(b_{ij}^\Delta, R, \sigma)$ -- to obtain a turbulence closure proper.

The combination of the correction terms and classifier is straightforward, namely the correction terms are replaced by the correction term multiplied by the classifier everywhere where they occur. For example for this constellation $(b_{ij}^\Delta, R, \sigma)$, it would be

\begin{equation}\label{eq:classapplied}
b_{ij}^\Delta \rightarrow \sigma \cdot b_{ij}^\Delta, \ \
R \rightarrow \sigma \cdot R
\end{equation}

\subsection{Flow field with optimal correction terms}\label{sec:optimal}
\begin{figure}[h!]
\centering
\includegraphics[clip,trim={0.\textwidth} {0.05\textwidth} {0.\textwidth} {0.02\textwidth},width=\textwidth]{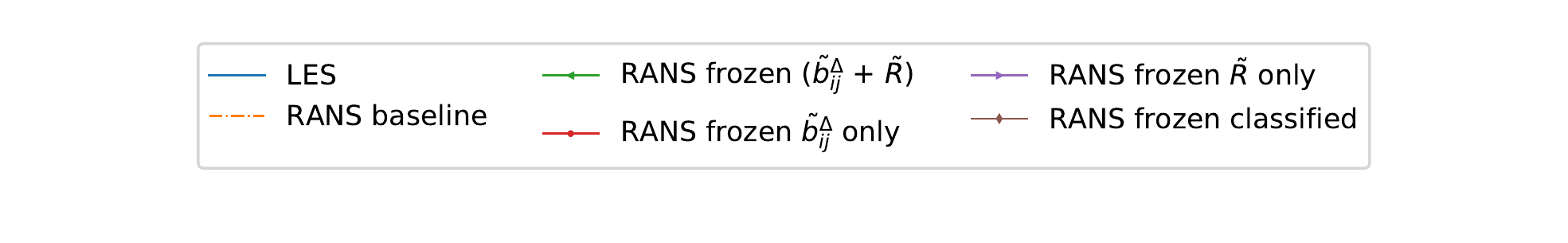} 
\includegraphics[clip,trim={0.0} {0.} {0.} {0.},width=\textwidth]{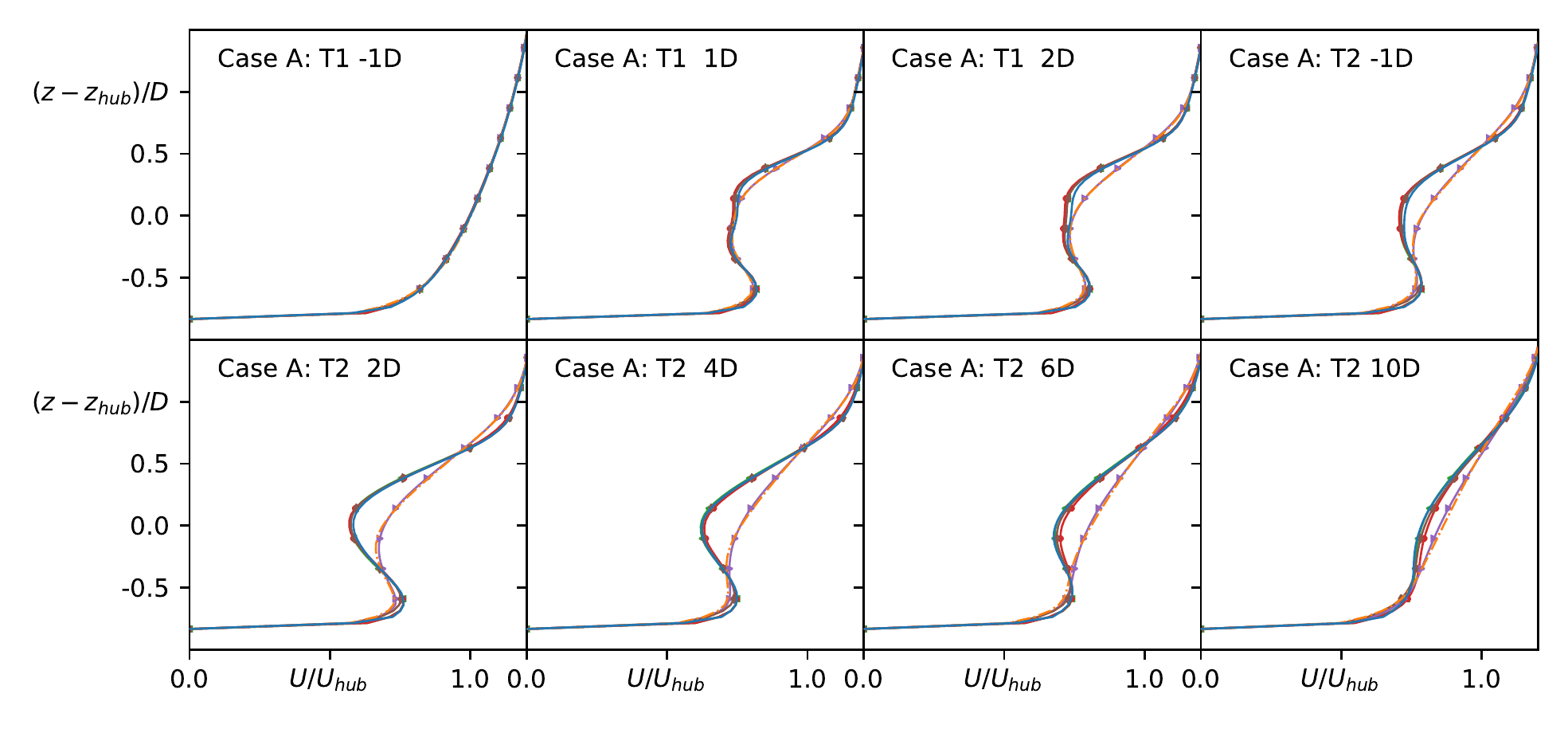} 
\includegraphics[clip,trim={0.0\textwidth} {0.} {0.0} {0.},width=\textwidth]{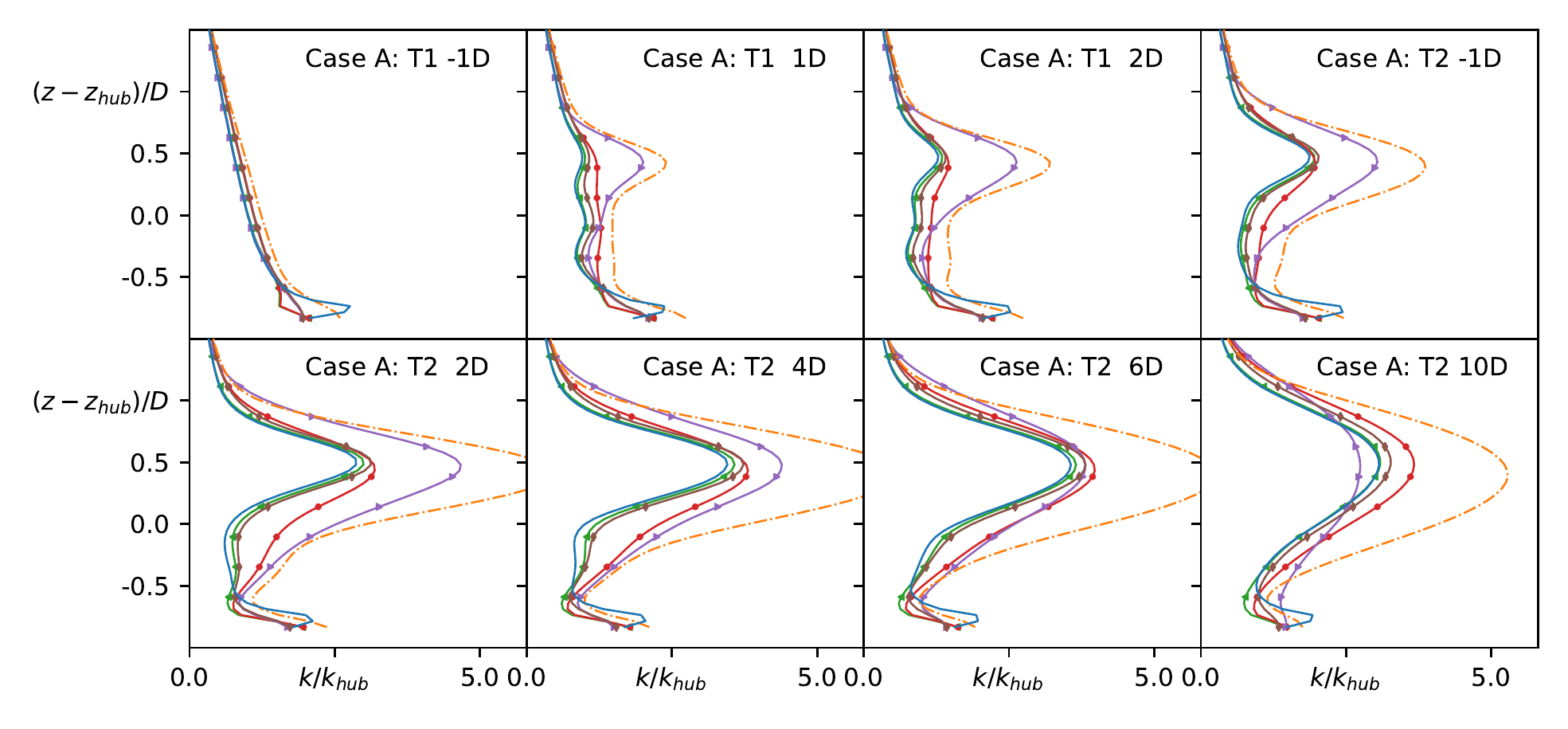} 
\caption{\label{fig:modelFrozenUk} Comparison between LES, RANS baseline, and frozen RANS with selective inclusion of the different components of the frozen correction terms as well as the ideal classifier.  Vertical slices of the velocity and TKE field up and downstream of the rotor plane of the two turbines of case A.}
\end{figure}

The optimal correction terms are derived for the three cases in the dataset using the frozen approach of Section~\ref{s:frozen}.  Subsequently, these optimal corrections are injected into the RANS simulation for all three test cases.  Figure \ref{fig:modelFrozenUk} shows the evolution of the flow velocity and the turbulent kinetic energy (TKE) (non-dimensionalised by their values at the turbine hub) as a function of non-dimensional height, at different stream-wise locations in the domain (from a distance of -1D upstream of the first turbine T1 to a distance of 10D downstream of the second turbine T2) for case A. These wake profiles are shown for the LES, the baseline RANS, and the frozen RANS simulations.

Optimally corrected RANS represent the best-case scenario that can be obtained using our methodology.  In the next subsection, the generalized models for the correction terms will introduce additional errors. The results in the figure show that indeed the optimal correction terms lead to an almost perfect match between the LES mean and frozen RANS velocity, as well as TKE fields.

We can see the effect of the wake classifier by multiplying the optimal corrections by the optimal classifier (as in \eqref{eq:classapplied}) - thereby restricting the regions in which these corrections are applied.  Again this leads to identical and close to identical profiles for the velocity and the TKE respectively.  This validates the criteria that were used for selecting the optimal classifier, and suggests its use is justified.

Figure \ref{fig:modelFrozenUk} also shows the relative importance of various correction terms for the prediction of the velocity and turbulent kinetic energy field.

The anisotropy correction term $\tilde{b}_{ij}^\Delta$ is significantly more important than the $k$-production correction term $\tilde{R}$.  In fact, if only a correct prediction of the velocity field is necessary, then the scalar term $\tilde{R}$ can be neglected completely. However, the scalar correction term $\tilde{R}$ does yield a significant improvement in the prediction of the turbulent kinetic energy over the case where only the tensor correction term $\tilde{b}_{ij}^\Delta$ is used.

\subsection{Training of correction terms and classifiers}\label{sec:learning}
In the following, the inputs and outputs of the training procedure for the correction terms without the classifier (Section~\ref{sec:modelWOcl}), the classifier itself (Section~\ref{sec:modelCL}), and the models with the classifier (Section~\ref{sec:modelWcl}) are described.

\subsubsection{Training of a correction model without classifier (Reference)}\label{sec:modelWOcl}

\begin{figure}[h!]
    \begin{minipage}{.5\textwidth}
        \centering
        \includegraphics[width=0.8\textwidth]{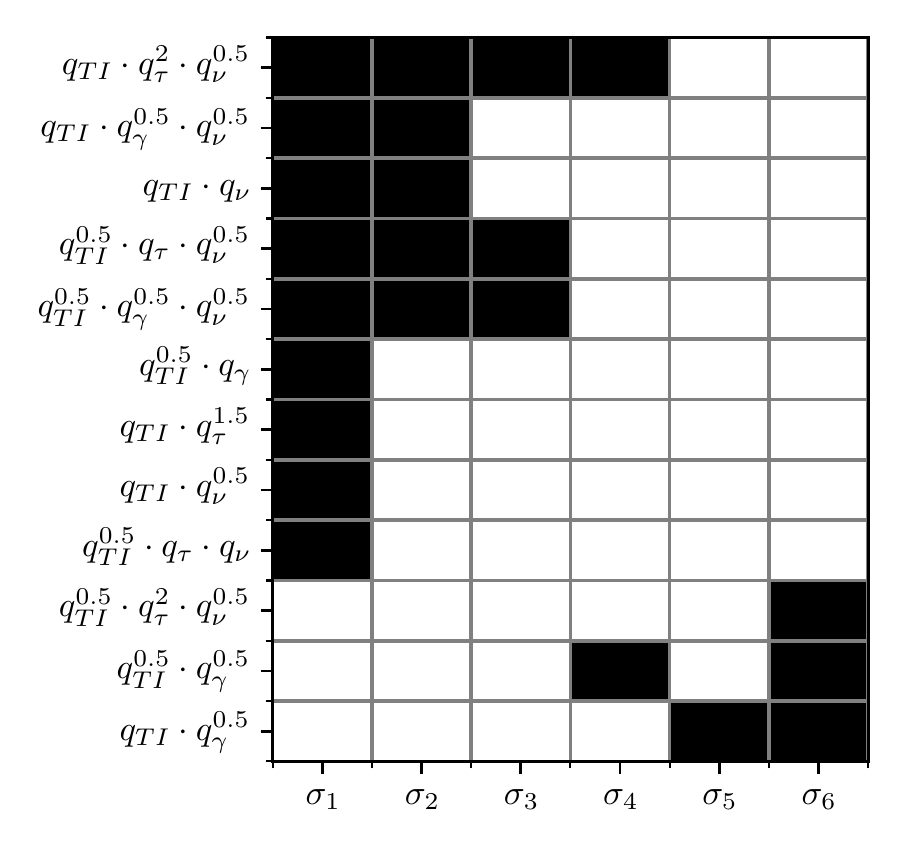} 
        \\
        \centering
        \includegraphics[width=\textwidth]{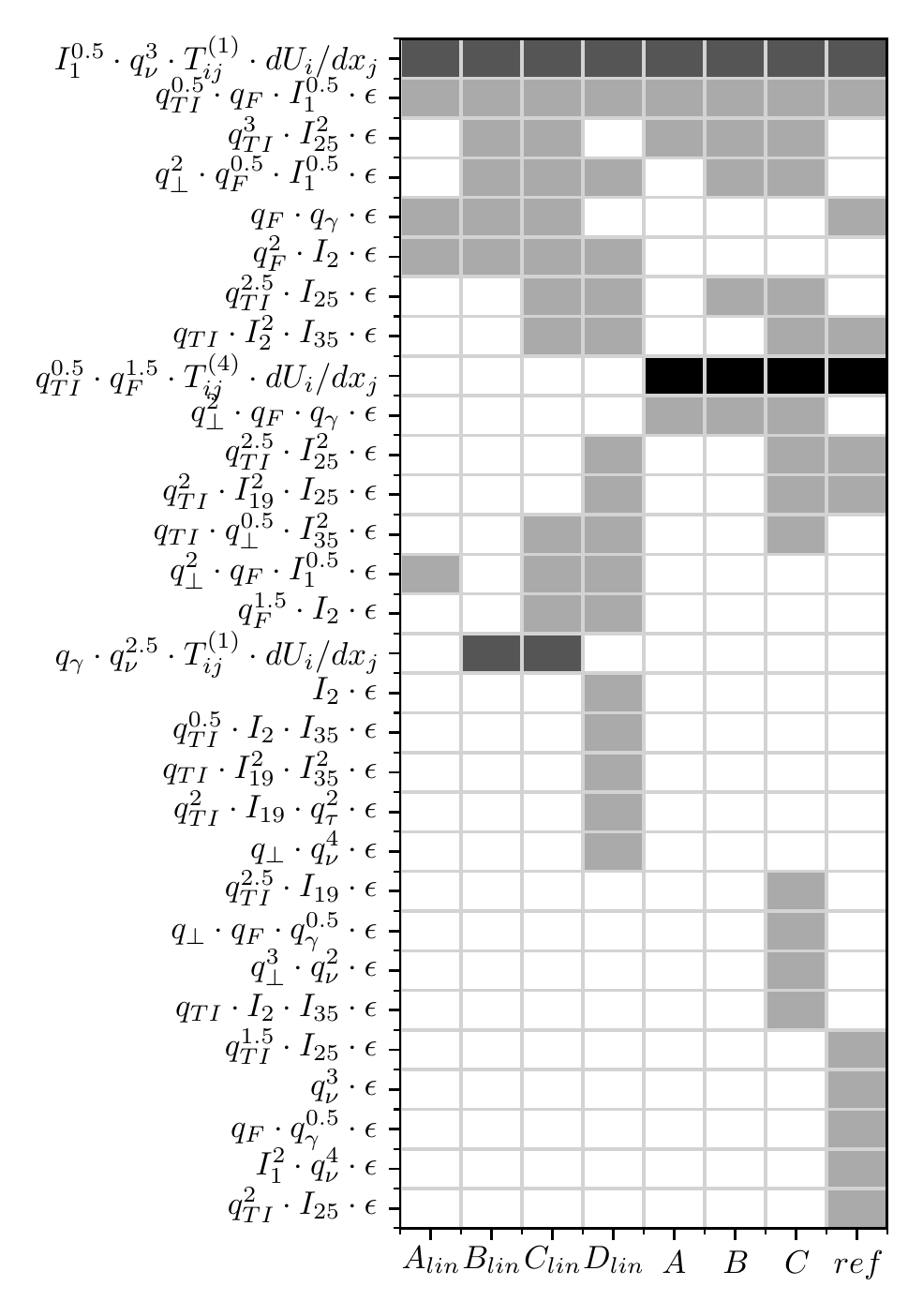} 
    \end{minipage}
    \begin{minipage}{.5\textwidth}
        \centering
        \includegraphics[width=\textwidth]{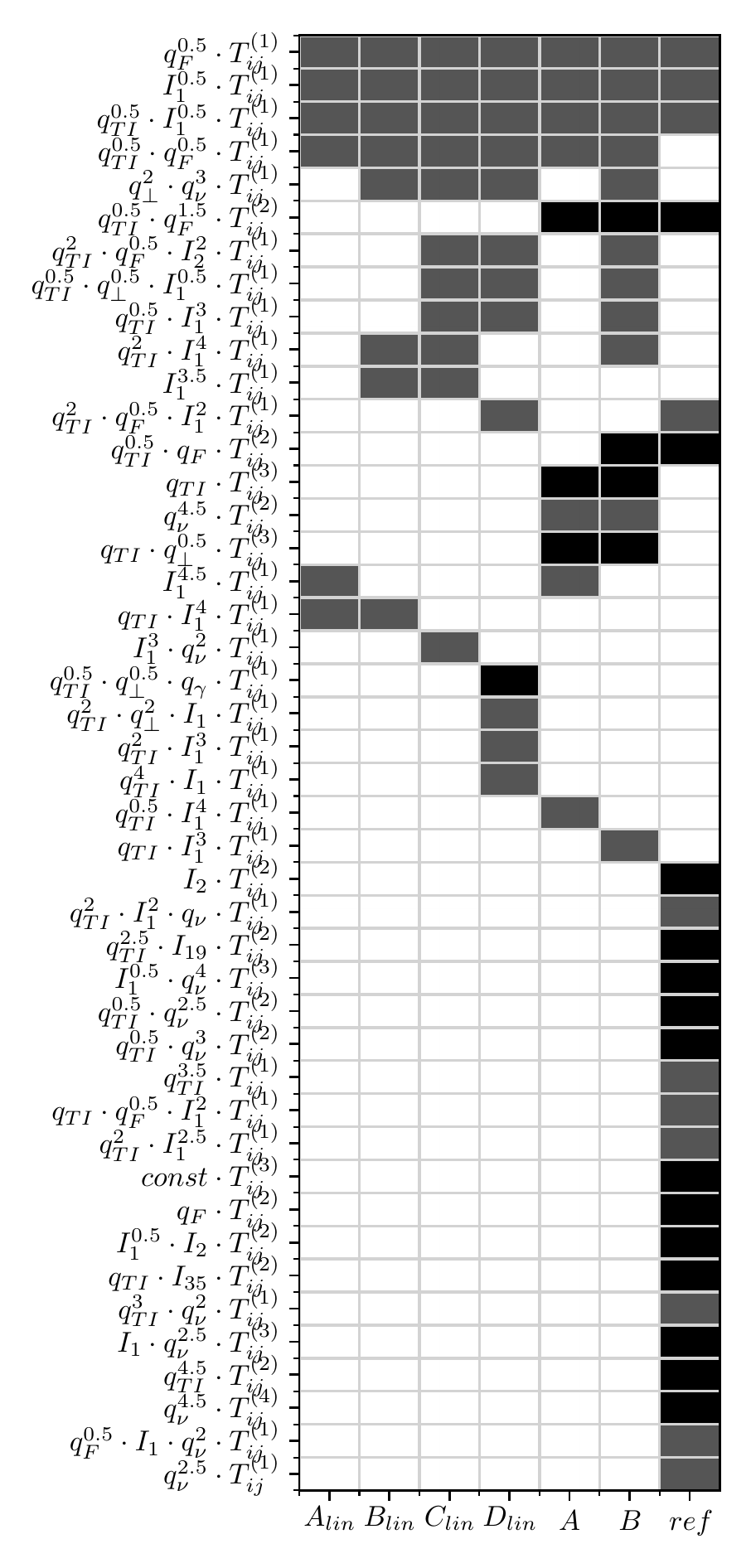} 
    \end{minipage}
    \caption{Visualization of the model terms for both correction terms and the classifier. The reference model from the publication without the classifier is also shown \cite{steiner2020b}. The colors pertain to terms which make no use of a tensor (light gray), are based on the shear strain rate tensor (gray) and make use of tensors other than the shear strain rate tensor (black). The figures pertain to (a,top left) the classifier $\sigma$, (b,bottom left) the TKE production correction $R$, and (c,right) the anisotropy correction $b_{ij}^\Delta$.}
    \label{fig:modelTerms}
\end{figure}

A model without a classifier was derived in \cite{steiner2020b} using the same elastic-net methodology as used in this paper. To evaluate whether the addition of a classifier is beneficial, this model will be used as a baseline here. Its accuracy, complexity and robustness will be compared with new models that do utilize a classifier. Figure \ref{fig:modelTerms} shows the terms of this correction model, marked as "ref".

\subsubsection{Training of the classifier}\label{sec:modelCL}

To model the classifier a smaller input feature set was used than for the correction terms, because satifactory results were obtained with a small input feature set. Hence, further extension of the feature set was not necessary.  Specifically, Table \ref{tab:clFeatures} shows the only first four features used (the remaining two features are used later in the correction terms).

\newcolumntype{b}{X}
\newcolumntype{s}{>{\hsize=.8\hsize\centering\arraybackslash}X}
\newcolumntype{x}{>{\hsize=.2\hsize\centering\arraybackslash}X}

\begin{table}[!htbp]
    \centering
	\begin{tabularx}{0.95\linewidth}{x b s s}
    \toprule \midrule
        ID & Description & Raw feature & Normalization\\ \midrule
  $q_{\gamma}$ & Shear parameter & $\left\|\frac{\partial U_i}{\partial x_j}\right\|$
  		& $\frac{\varepsilon}{k}$ \\
  $q_{\tau}$ & Ratio of total to normal Reynolds stresses &  $||\overline{u_i'u_j'}_{Boussinesq}||$    &  $k$ \\
  $q_{\nu}$   &   Viscosity ratio &   $\nu_t$    &   $100\nu$ \\
  $q_{TI}^{{\dagger}}$   &   Turbulence intensity  &   $k$    &   $\frac{1}{2}U_iU_i$ \\
  $q_{F}^{{\dagger}}$ & Actuator forcing
  		& $\left\|F_{cell}\right\|$ & $\frac{1}{2} \rho_0 A_{cell} \left\| U \right\|^2$ \\
    $q_{\perp}^{{\dagger}}$ & Nonorthogonality between velocity and its gradient
    & $|U_iU_j \frac{\partial U_i}{\partial x_j}|$
    & $\sqrt{U_lU_lU_i\frac{\partial U_i}{\partial x_j}U_k \frac{\partial U_k}{\partial x_j}}$ \\
	\midrule \bottomrule
	\end{tabularx}
    \caption{List of non-dimensionalized physical features used in the model discovery phase and their precise definition. The features that are not Galilean invariant are marked with ${\dagger}$.}
    \label{tab:clFeatures}
\end{table}

Varying the regularization parameters of the elastic net resulted in the identification of a large number of classifiers, of which six were selected for further testing.
The complexity of the chosen classifiers ranges from one to nine terms; and notably more complex models did not show a significant increase in accuracy during training.  Figure \ref{fig:modelTerms}(a) visualises the terms used.  There is significant overlap between the terms used by the classifiers, notably features of turbulence intensity, velocity shear and eddy viscosity ratio are dominant in all.

\begin{figure}[h!]
\centering
\includegraphics[clip,trim={0.1\textwidth} {0.0\textwidth} {0.1\textwidth} {0.03\textwidth},width=0.95\textwidth]{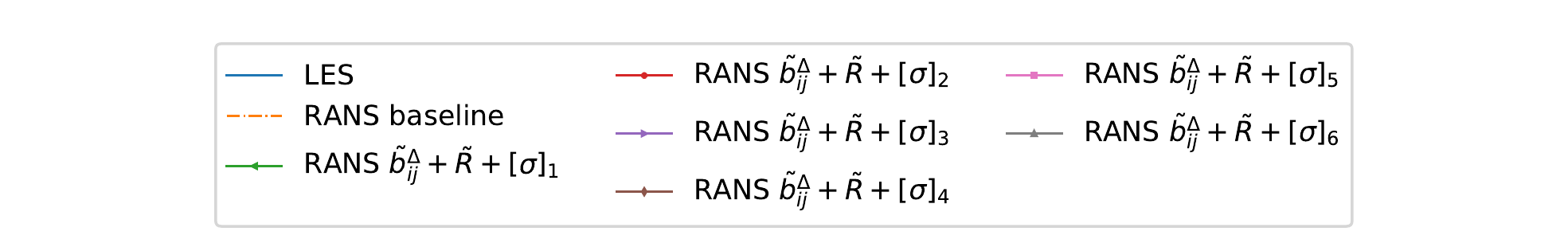} 
\includegraphics[clip,trim={0.0\textwidth} {0.} {0.\textwidth} {0.},width=0.95\textwidth]{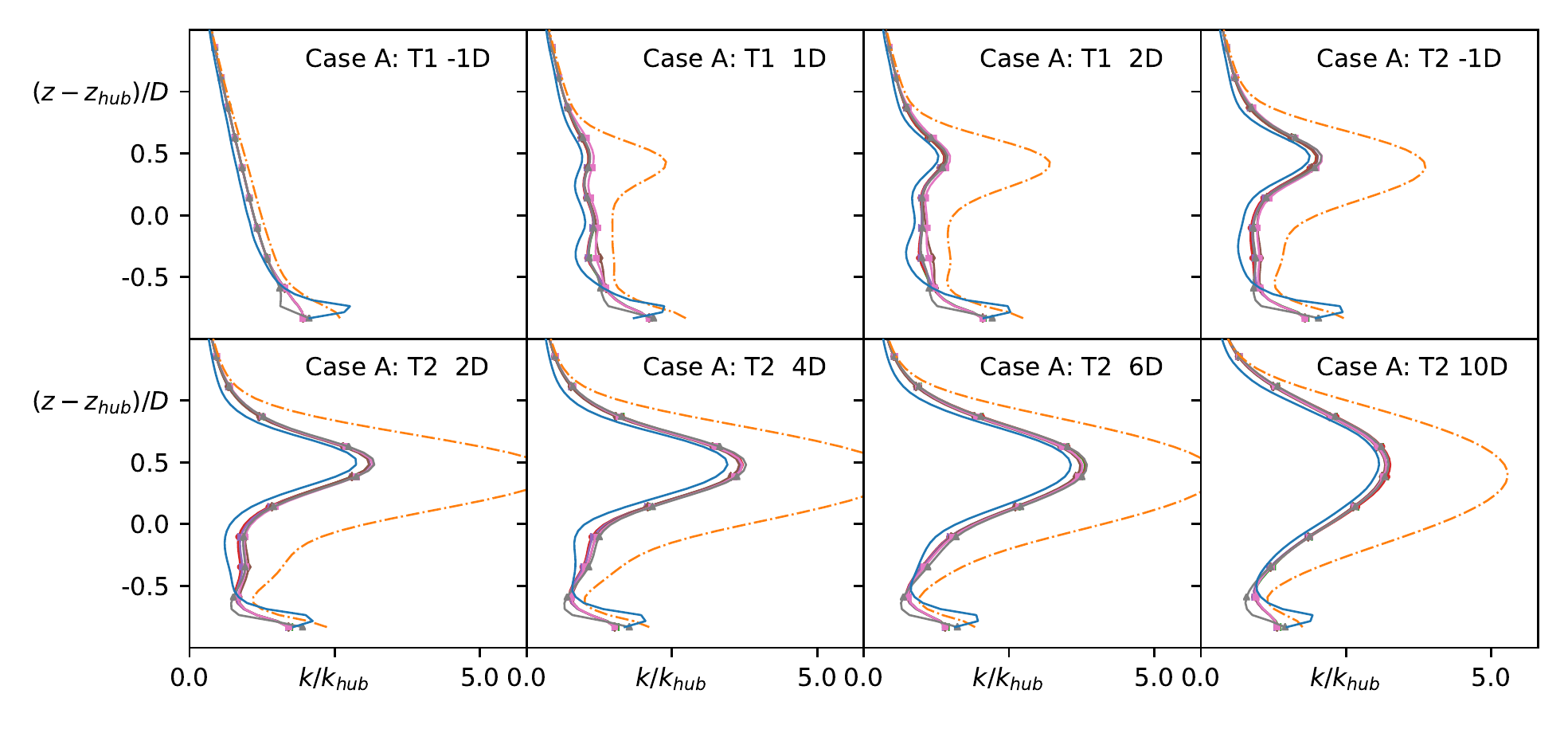} 
\includegraphics[clip,trim={0.0\textwidth} {0.} {0.\textwidth} {0.},width=0.95\textwidth]{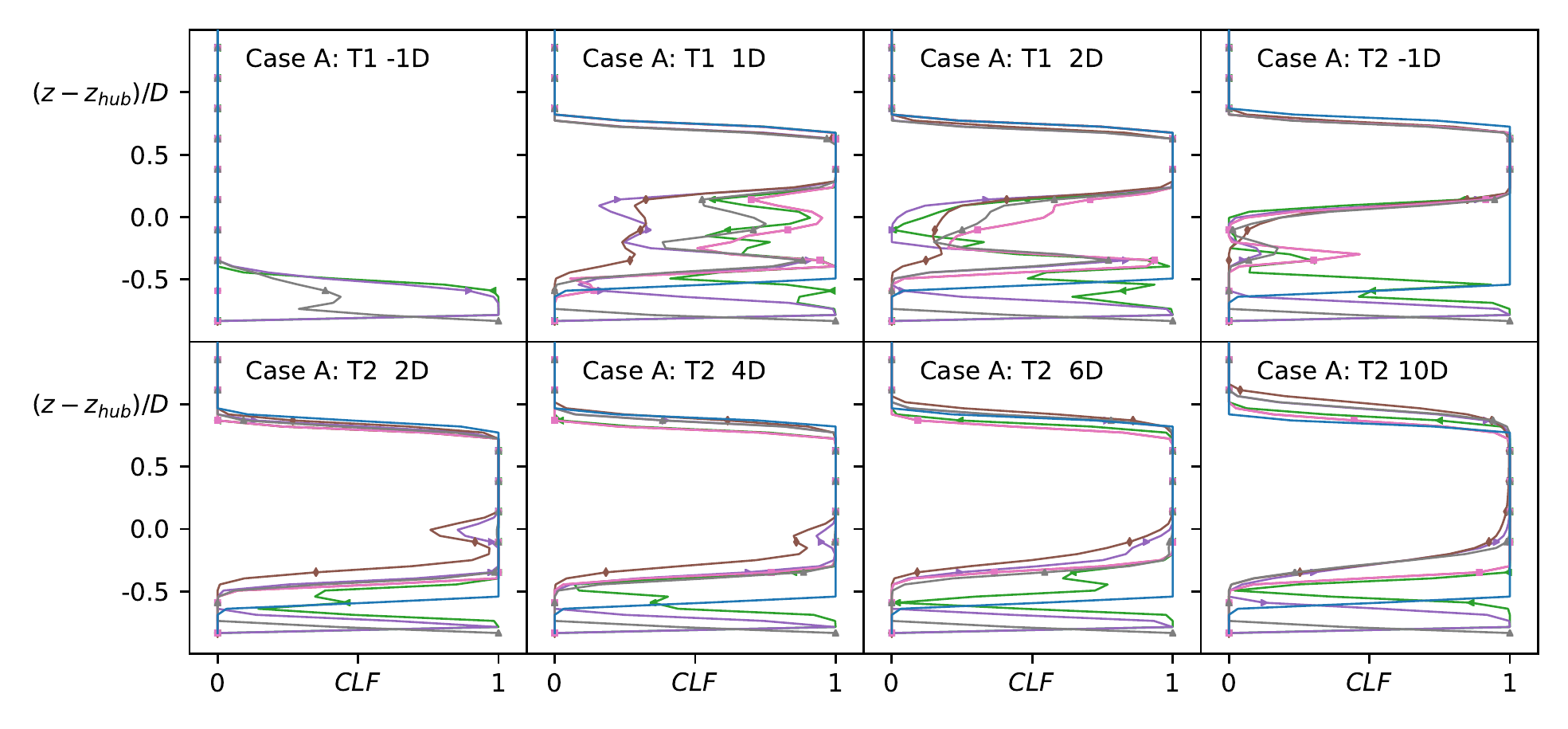} 
\caption{Comparison between LES, RANS baseline, and corrected RANS simulations with frozen correction terms and coupled classifiers. The figures depict vertical slices the TKE (top) and classifier (bottom) field up and downstream of the rotor plane of the two turbines of case A.}
\label{fig:caseAclfOnlyUk}
\end{figure}

All six were implemented in the RANS solver in combination with the optimal correction terms, i.e.\  $(\tilde{b}_{ij}^\Delta, \tilde{R}, \sigma_k)$ for $k\in\{1,\dots,6\}$.  Figure \ref{fig:caseAclfOnlyUk} shows the vertical distribution of the turbulent kinetic energy (top) and the classifier fields (bottom) for the different classifiers. The velocity profiles are essentially unaffected by the choice of classifier and are not shown in the figure. Furthermore, there is minimal variation in the TKE profiles, except close to the wall. The classifier values themselves show quite some spread in the bottom part of the wake and towards the wall which does not seem to affect the velocity and TKE fields much. There are two reasons for this. Firstly, the corrections are generally small in the lower part of the wake; and secondly, the blending function ensures that the corrections go towards zero once the wall is approached.

On this basis only classifiers $\left[\sigma\right]_3$, $\left[\sigma\right]_5$, $\left[\sigma\right]_6$ will be further investigated.  Classifier $\left[\sigma\right]_5$ is chosen because it is the simplest. Classifier $\left[\sigma\right]_3$ and $\left[\sigma\right]_6$ are chosen because they are a bit more complex and show a different near-wall behaviour.

\subsubsection{Training of models with a classifier}
\label{sec:modelWcl}

Finally we train the models that will be combined with a classifier when coupled to the flow solver.  The classifier is always trained first (see previous section), and used to discard points from the training data-set.  As such the correction model only has to reproduce the optimal correction where the classifier is active, potentially allowing for a better fit with simpler models.

\begin{table}
  \centering
   \begin{tabular}{cc}
   \toprule \midrule
   Invariant ID & Definition \\
   \midrule
   $I_1$ & $\textbf{S}^2$  \\
   $I_2$ & $\boldsymbol{\Omega}^2$  \\
   $I_{19}$ & $\boldsymbol{\Omega} \textbf{A}_k \textbf{S}^2$  \\
   $I_{25}$ & $\textbf{A}_k^2 \textbf{S} \boldsymbol{\Omega} \textbf{S}^2$  \\
   $I_{35}$ & $\textbf{A}_p \textbf{A}_k \textbf{S}^2$  \\
   \midrule \bottomrule
   \end{tabular}
   \qquad
  \begin{tabular}{ccc}
  \toprule \midrule
  Tensor ID & Definition & Normalization \\
  \midrule
  $\textbf{S}$
    & $\frac{1}{2} \left( \frac{\partial u_i}{\partial x_j}+\frac{\partial u_j}{\partial x_i}\right)$
    & $\frac{\varepsilon}{k}$\\
  $\boldsymbol{\Omega}$
    &  $\frac{1}{2} \left( \frac{\partial u_i}{\partial x_j}-\frac{\partial u_j}{\partial x_i}\right)$
    & $\frac{\varepsilon}{k}$\\
  $\textbf{A}_k$
    & $-I \times \nabla p$
    & $\frac{\varepsilon}{\sqrt{k}}$ \\
  $\textbf{A}_p$
    & $-I \times \nabla k$
    & $\rho_0 \left\| u \nabla u \right\|$\\
  \midrule \bottomrule
  \end{tabular}
  \caption{List of invariants used in the model discovery phase and their precise definition.}
  \label{tab:invariants}
\end{table}

For both the anisotropy and the $k$-production correction terms, two separate model discovery runs were performed: I.\  using only the strain rate tensor, and II.\  using the first four tensors of Pope's integrity basis.  As a consequence, the resulting models are either linear (I) or nonlinear (II) eddy viscosity models.  Linear eddy viscosity models can be written

\begin{equation}
    \label{eq:psiDef}
    b_{ij}^\Delta = \psi(\cdot) \cdot T_{ij}^{(1)} = \psi(\cdot) \frac{k}{\epsilon} S_{ij},
\end{equation}

for some scalar-valued function $\psi(\cdot)$, and are preferable for their simplicity and stability, though they are less likely to be able to reproduce the flow-field in these cases where turbulence anisotropy is demonstrably important.

Table \ref{tab:clFeatures} and Table \ref{tab:invariants} present the input feature set consisted of the physical parameters and the invariants, respectively.  From the resulting model sets, the selection procedure described in Section \ref{s:sparsereg} was used to choose a handful of models.  Initial tests were performed using an partially coupled approach where the correction term in question was coupled to the flow solver while the other correction term and the classifier was used in optimal form. The models were selected on basis of their complexity: initially the simplest available model was examined and then the complexity was increased until the models either stopped converging or stopped yielding an improvement over the previous model.

Figure \ref{fig:modelTerms} shows the different terms of the chosen model formulations for the anisotropy correction term. The terminology for the linear and the nonlinear eddy viscosity models is $\left(A_{lin},B_{lin},C_{lin},D_{lin}\right)$ and $\left(A,B\right)$, respectively. The model complexity ranges from 6 to 15 terms. There are four terms that are used by all models, they all contain the strain rate tensor $S_{ij}$ combined with the two physical features $q_{TI}$ and $q_F$, and the invariant $I_1$. In the remaining terms, the physical features $q_\nu$ and $q_\perp$ are highly represented, followed by the invariant $I_2$. The most frequently used feature is $I_1$. For the nonlinear eddy viscosity models a large overlap between the terms with nonlinear tensors is present. It is interesting to note that these terms do not involve the first invariant $I_1$.

\begin{figure}[h!]
\centering
\includegraphics[clip,trim={0.0\textwidth} {0.} {0.\textwidth} {0.},width=\textwidth]{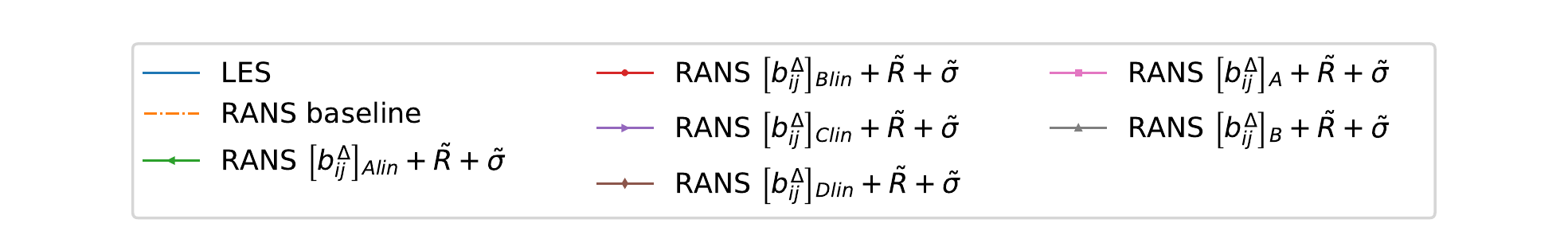} 
\includegraphics[clip,trim={0.0\textwidth} {0.} {0.\textwidth} {0.},width=\textwidth]{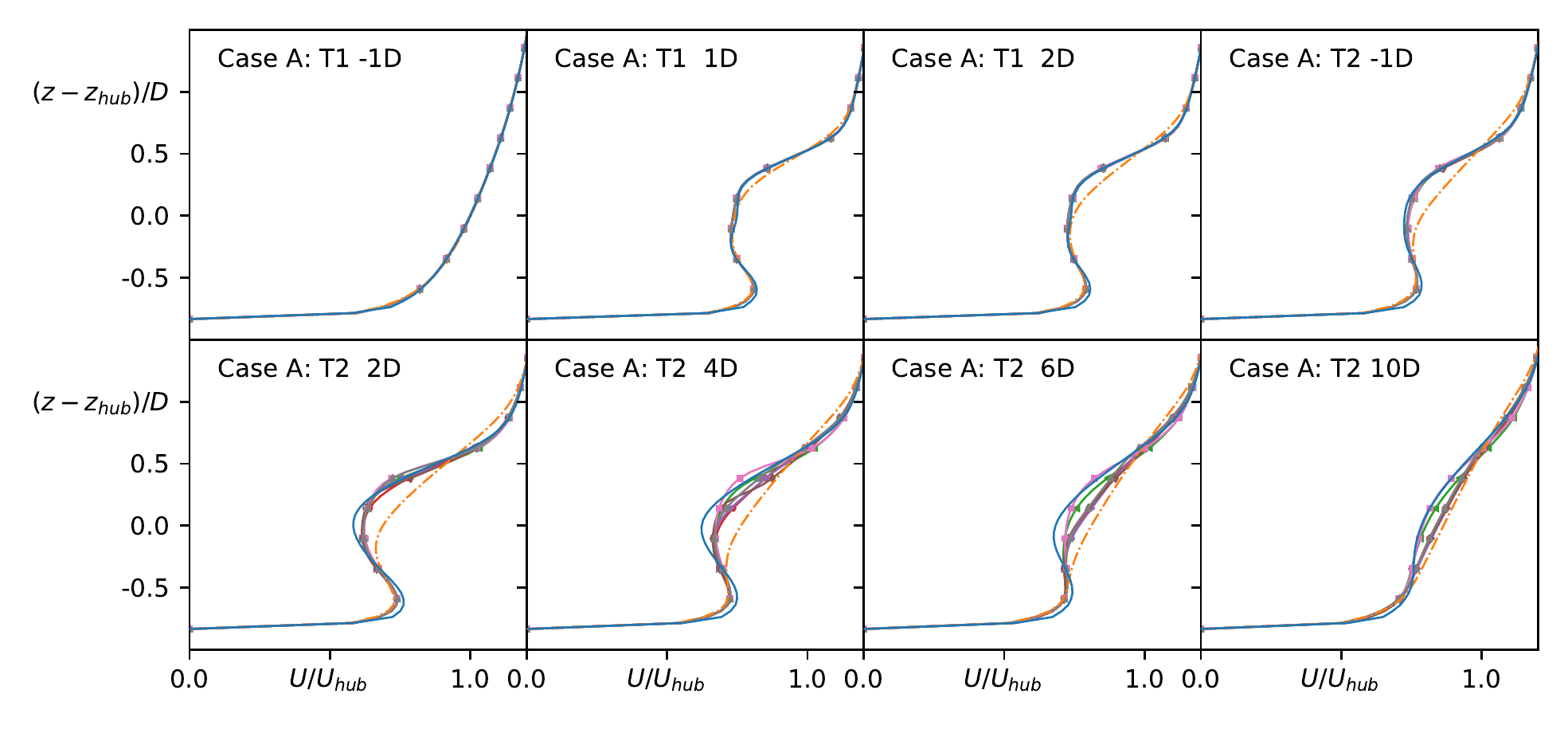} 
\includegraphics[clip,trim={0.0\textwidth} {0.} {0.\textwidth} {0.},width=\textwidth]{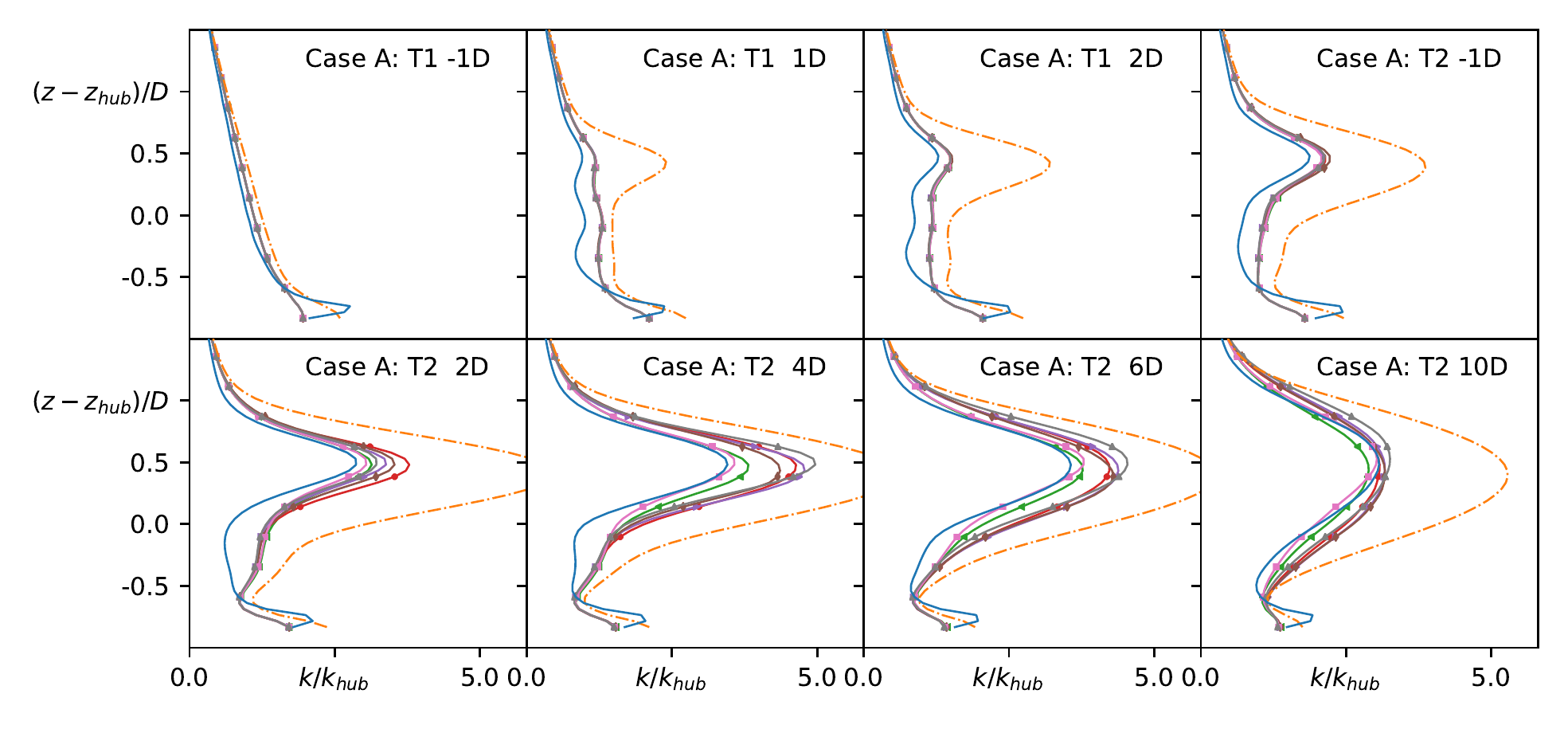} 
\caption{Comparison between LES, RANS baseline, and corrected RANS coupled with different formulations for the anisotropy correction $b_{ij}^\Delta$. The TKE production correction $R$ and the classifier $\sigma$ are frozen. Vertical slices of the velocity (top) and TKE (bottom) field up and downstream of the rotor plane of the two turbines of case A are shown.}
\label{fig:caseARfixedUk}
\end{figure}

Figure~\ref{fig:caseARfixedUk} shows the results of the partially coupled runs with different anisotropy corrections for case A. Again, the spread between the different models is larger for the $k$ profiles than for the velocity profiles. Further, there is no significant spread between the models for the first turbine's wake. However, the models differ for the second turbine. Overall, the simplest linear and the simplest nonlinear model yield the most consistent improvement over the baseline model.

Moving on to the TKE correction term $R$, again the distinction between linear and nonlinear models is made. Figure \ref{fig:modelTerms} shows a comparison of the selected models based on their mathematical formulation. The first observation is that most of the terms are source terms and all of the models have between one and three production terms. Two terms are used in all the different corrections: (i) a production term and (ii) a source term related to the actuator forcing. When compared to the anisotropy correction models, the TKE production correction models use a wider range of features, most likely because this correction term is more complex and not as strongly related to velocity shear. Further,  the fourth tensor of the integrity basis is used in all the fully nonlinear models and the first invariant is not used in conjunction with this tensor.

\begin{figure}[h!]
\centering
\includegraphics[clip,trim={0.\textwidth} {0.} {0.\textwidth} {0.},width=\textwidth]{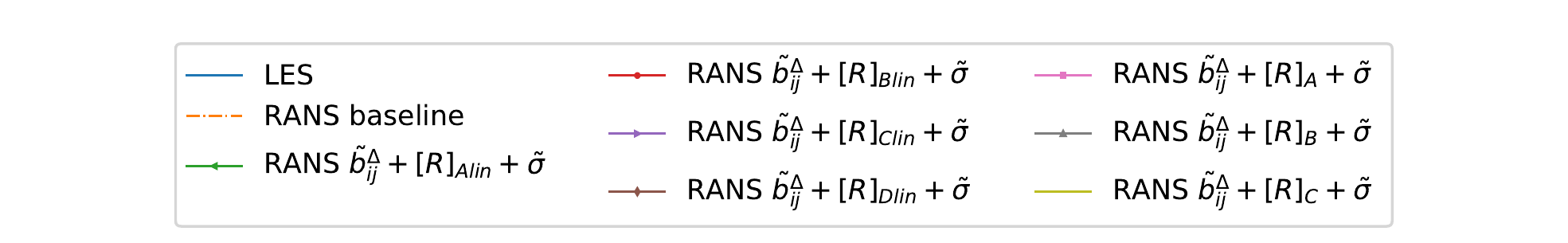} 
\includegraphics[clip,trim={0.\textwidth} {0.} {0.\textwidth} {0.},width=\textwidth]{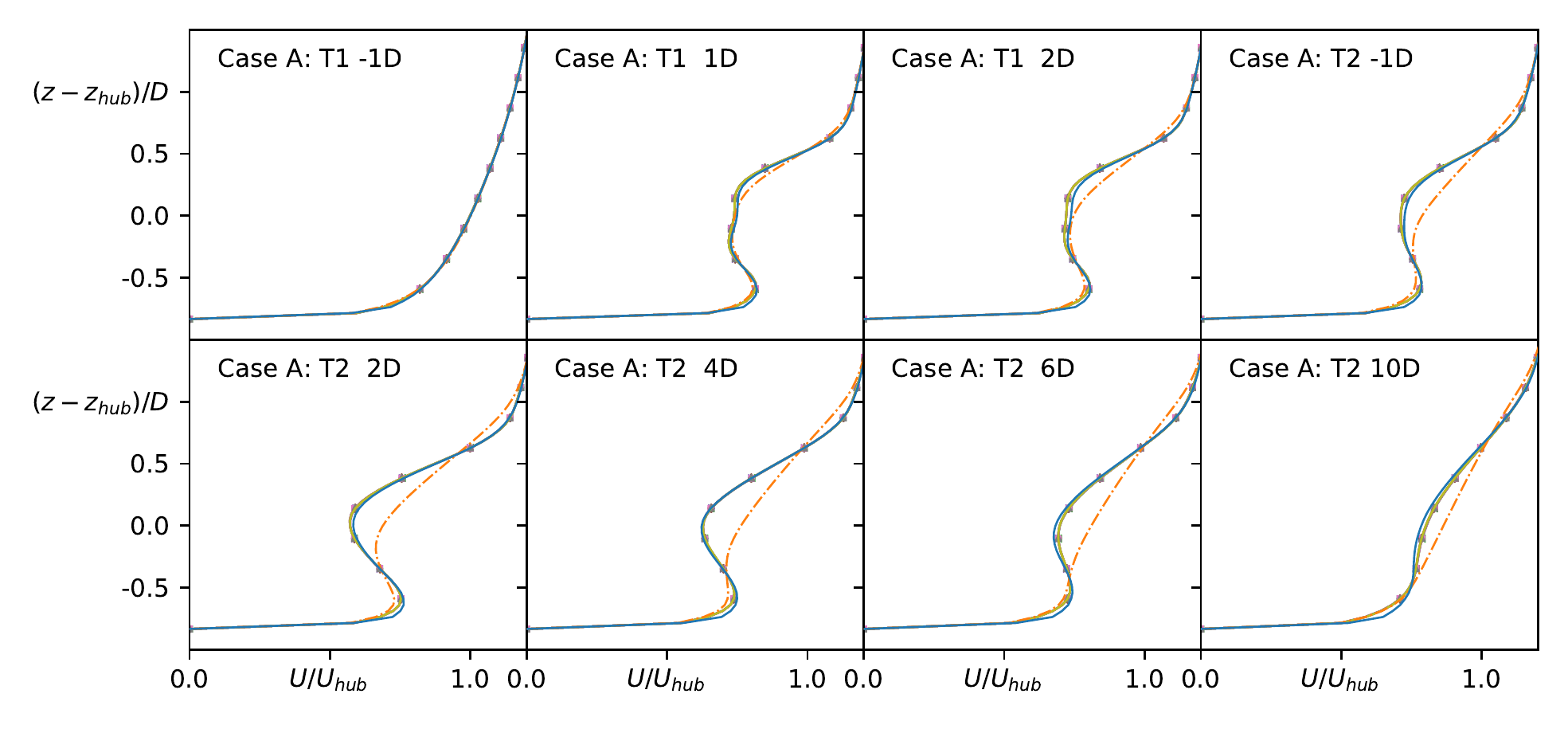} 
\includegraphics[clip,trim={0.\textwidth} {0.} {0.\textwidth} {0.},width=\textwidth]{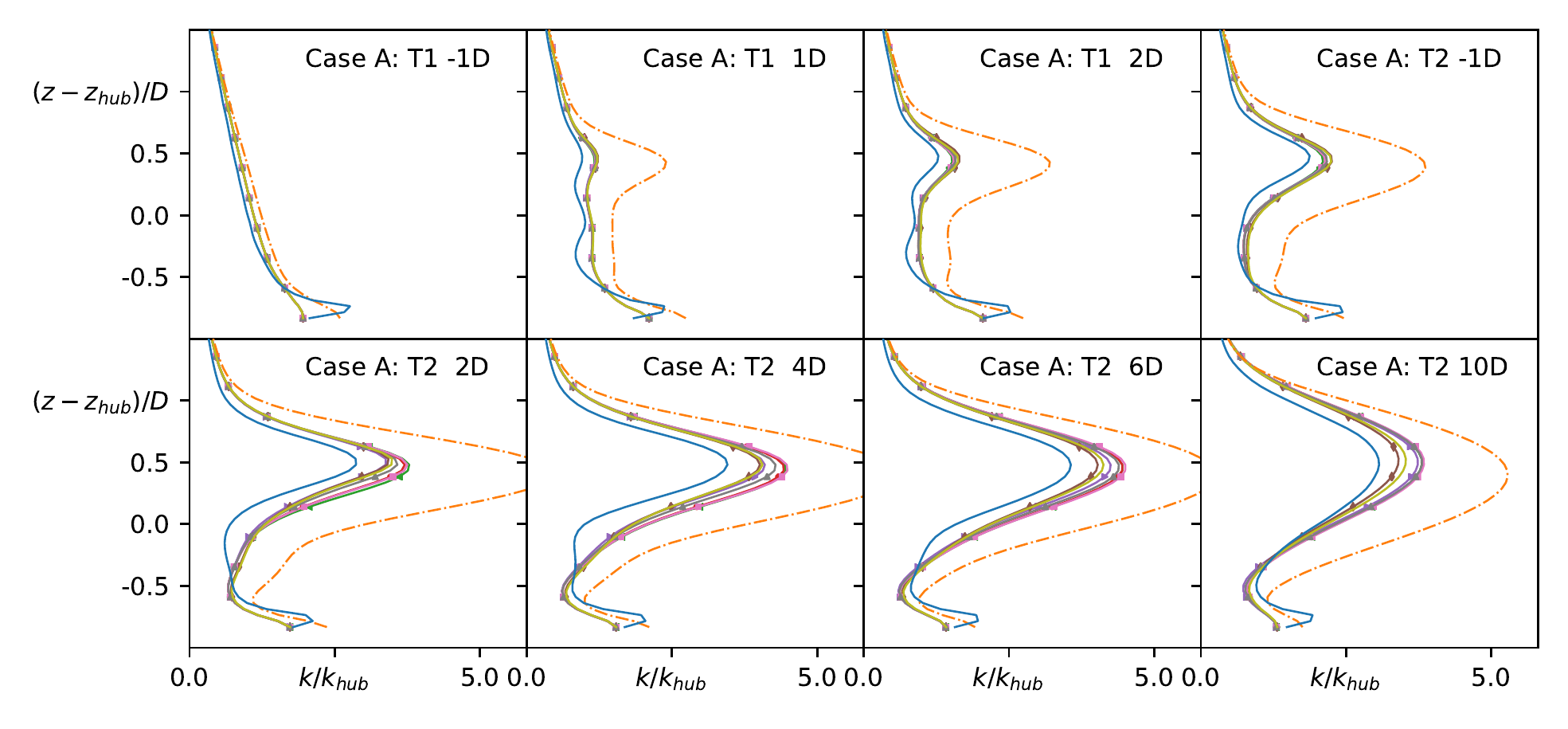} 
\caption{Comparison between LES, RANS baseline, and corrected RANS coupled with different formulations for the TKE production correction $R$. The anisotropy correction $b_{ij}^\Delta$ and the classifier $\sigma$ are frozen. Vertical slices of the velocity (top) and TKE (bottom) field up and downstream of the rotor plane of the two turbines of case A are shown.}
\label{fig:caseABfixedUk}
\end{figure}

Finally, Figure \ref{fig:caseABfixedUk} shows what happens when these models are coupled with the flow solver using the same partial coupling approach as for the anisotropy correction, so $\tilde{b}_{ij}^\Delta + R + \tilde{\sigma}$. Since this correction term mainly affects the turbulent kinetic energy, there is no visible difference in the velocity profiles. There is some spread in the turbulent kinetic energy profiles which is largest in the near wake of the second turbine. The most complex linear and the most complex nonlinear model yield the most consistent improvement over the baseline model, but the difference between the models is not significant.

\subsection{Robustness of correction terms}
\label{sec:robustness}

In this section, an analysis of the numerical stability of the correction terms is presented, important since we observe the introduction of the classifier makes the models a little more prone to instability.  The authors suspect this is because models derived in conjunction with classification are not required to be zero in non-wake regions. As such the models have become more sensitive to changes in the input features and tensors.  In the previous publication already a brief stability analysis was carried out where the focus was on the difference between fixed and coupled correction terms \cite{steiner2020b}.

Here numerical instability related to the correction terms is observed for some model combination and some cases. The instability manifests itself in two different forms in the correction terms. The anisotropy correction sometimes diverges in the near wake of the turbines and the TKE production correction sometimes diverges at the rotor disk. However, the authors think both manifestations are based on the same underlying instability. Two factors contribute to this:
\begin{enumerate}
    \item Since the baseline $k-\epsilon$ model tends to over-predict $k$ and eddy viscosity,
    and since the actuator disc model does not remove turbulence energy from the flow, our
    the correction terms act to remove energy almost everywhere in the flow domain; and
    \item In most of our models the dominant term(s) depend on products of the shear strain invariant $I_1$, the normalized shear stress tensor $T_{ij}^{(1)}$ and the velocity gradient tensor.
\end{enumerate}
So if locally one component of the velocity gradient tensor gets too large, the effect can be amplified by in the model (which can contain higher-powers of $\nabla U$), removing more energy, which increases velocity gradients further.

To break this positive coupling loop with minimal intervention, two limiters are proposed for the two manifestations of the instability:
\begin{itemize}
    \item \textbf{Eddy viscosity limiter:} Based on the $k$-$\varepsilon$-$f_P$ model \cite{laan2013} the linear components of our anisotropy models were limited to avoid removing too much energy from the simulation:
    \begin{equation}
     \psi = \min \left( 0.8 \cdot \frac{\varepsilon}{k^2} \cdot \nu_t,
     \psi \right) ,
    \end{equation}
    which corresponds to $f_P>0.2$. This threshold was derived based on an analysis of the available data-set and is chosen so low that it should only be active when there is indeed a positive coupling loop present.

    \item \textbf{Form error limiter:} This address tendency of the correction models to aggressively remove energy near the actuator discs.  This limiter is only active in areas where actuator forcing is applied, and is chosen based on the Boussinesq turbulent kinetic energy production as:
    \begin{equation}
        R = \text{sgn} \left(R\right) \cdot min
            \left(0.5 P_k^{Boussinesq},\left\lvert R \right\rvert\right).
    \end{equation}
\end{itemize}

\begin{figure}[h!]
\centering
\includegraphics[width=\textwidth]{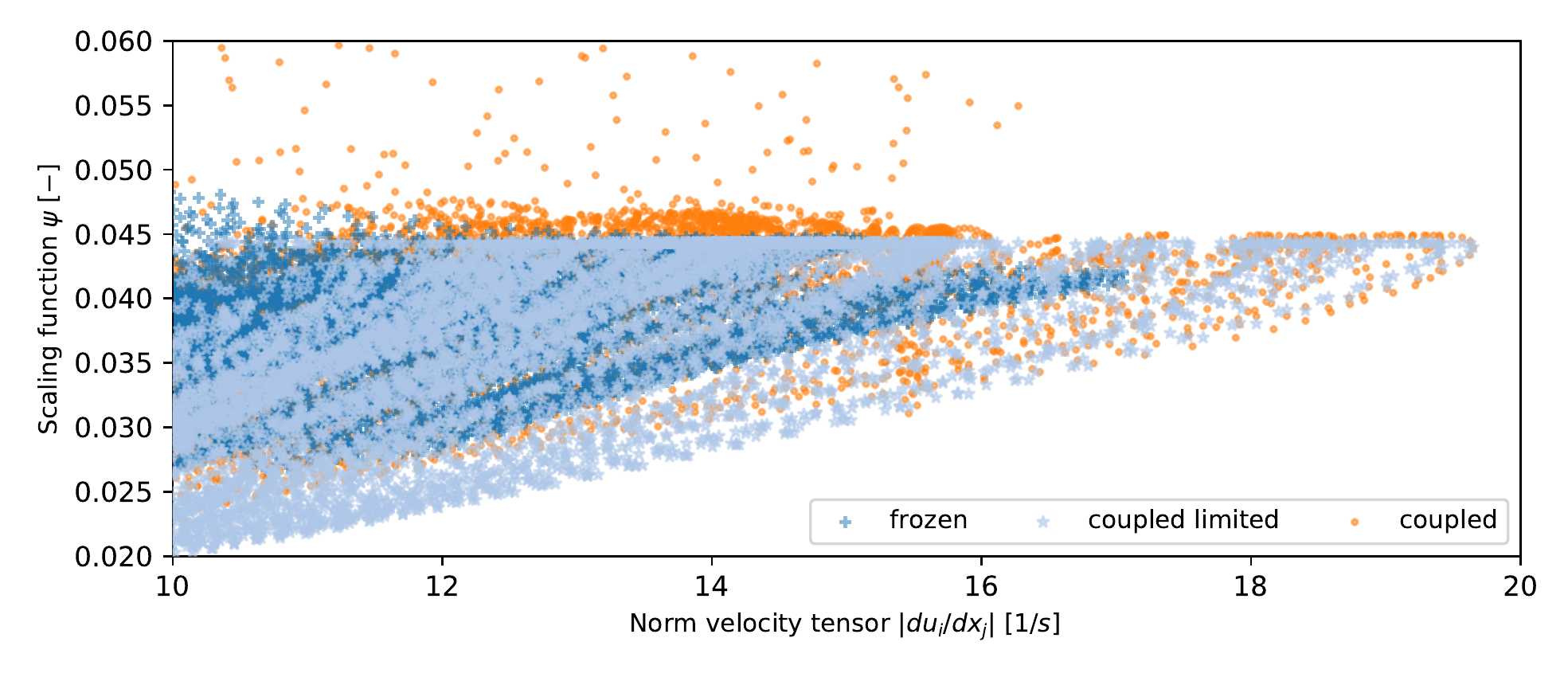} 
\includegraphics[width=\textwidth]{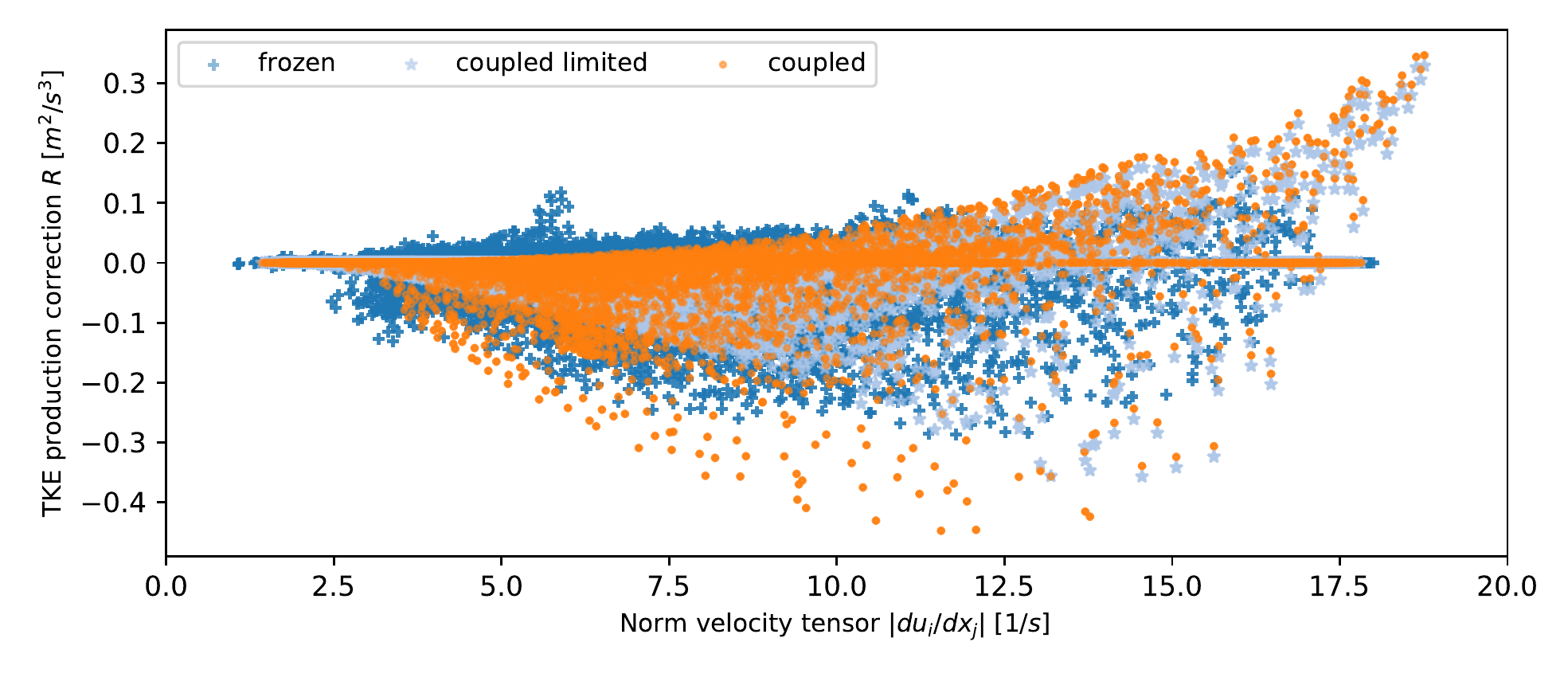} 
\caption{Scatterplot for the two correction terms for (top) the wake and (bottom) the points with non-zero actuator forcing of all three turbines of case C in combination with the two simplest correction models $\left[ b_{ij}^\Delta\right]_{Alim},\left[ R\right]_{Alim}$ and the most simple classifier $\left[ \sigma\right]_{5}$. The top plot shows the effect of the eddy viscosity limiter on the scaling function $\psi$. The bottom plot show the effect of the form error limiter on the correction term $R$.}
\label{fig:gradScatter}
\end{figure}

Figure \ref{fig:gradScatter} visualizes the correlation between velocity gradient and correction magnitude as well as the effect of the two limiters for the two simplest linear models. The top plot \ref{fig:gradScatter}(a) shows the scaling function for the anisotropy correction $\psi$. Equations \eqref{eq:bij} and \eqref{eq:psiDef} show that larger values of the scaling function $\psi$ lead to smaller eddy viscosity values, because the correction is subtracted from the Boussinesq expression for the eddy viscosity. One can see that the limiter effectively limits the maximum of said function - so the minimum of the eddy viscosity - and that in turn the magnitude of the correction does not only decrease where the limiter is active, but also in the cells close to it. The bottom plot \ref{fig:gradScatter}(b) shows the TKE production correction term in the wake of multiple turbines and one can see how the limiter effectively removes outliers.

\begin{figure}[h!]
\centering
\includegraphics[width=\textwidth]{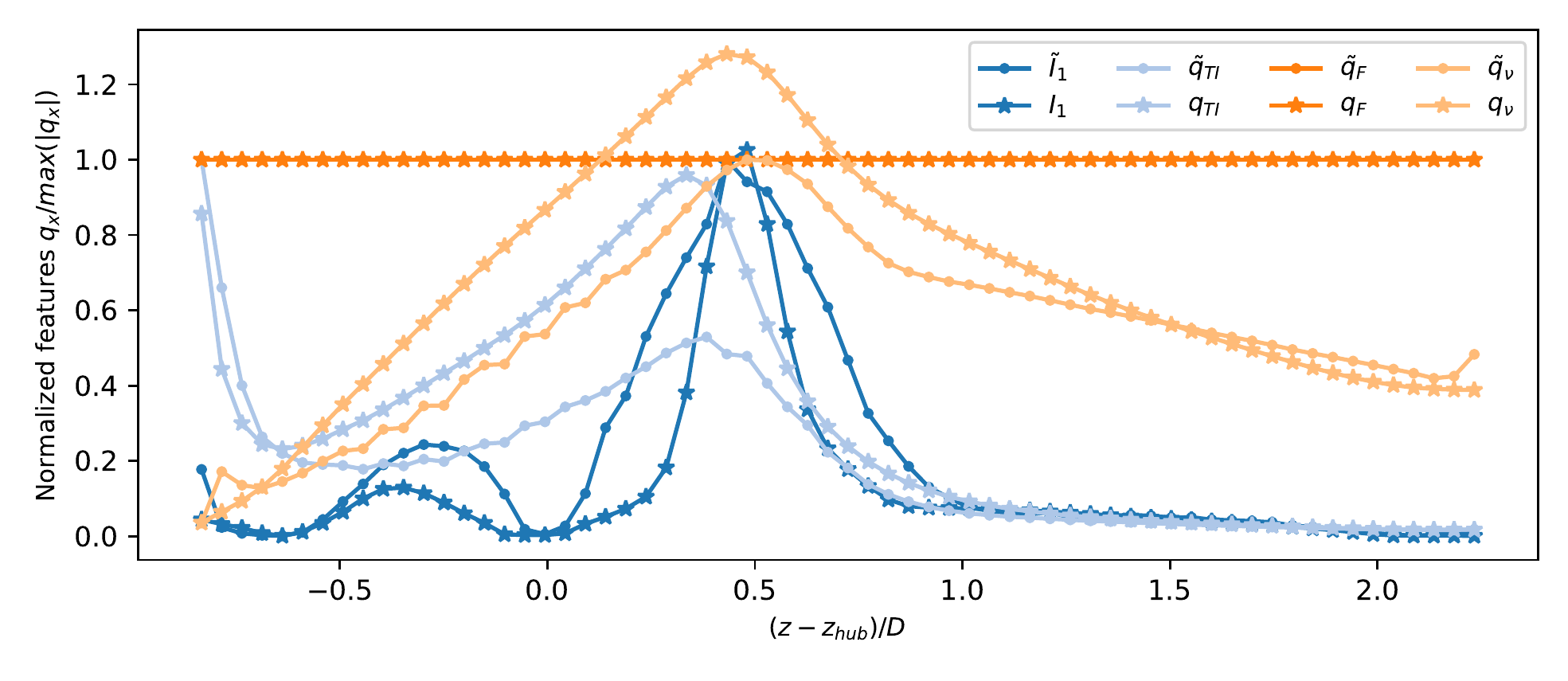} 
\includegraphics[width=\textwidth]{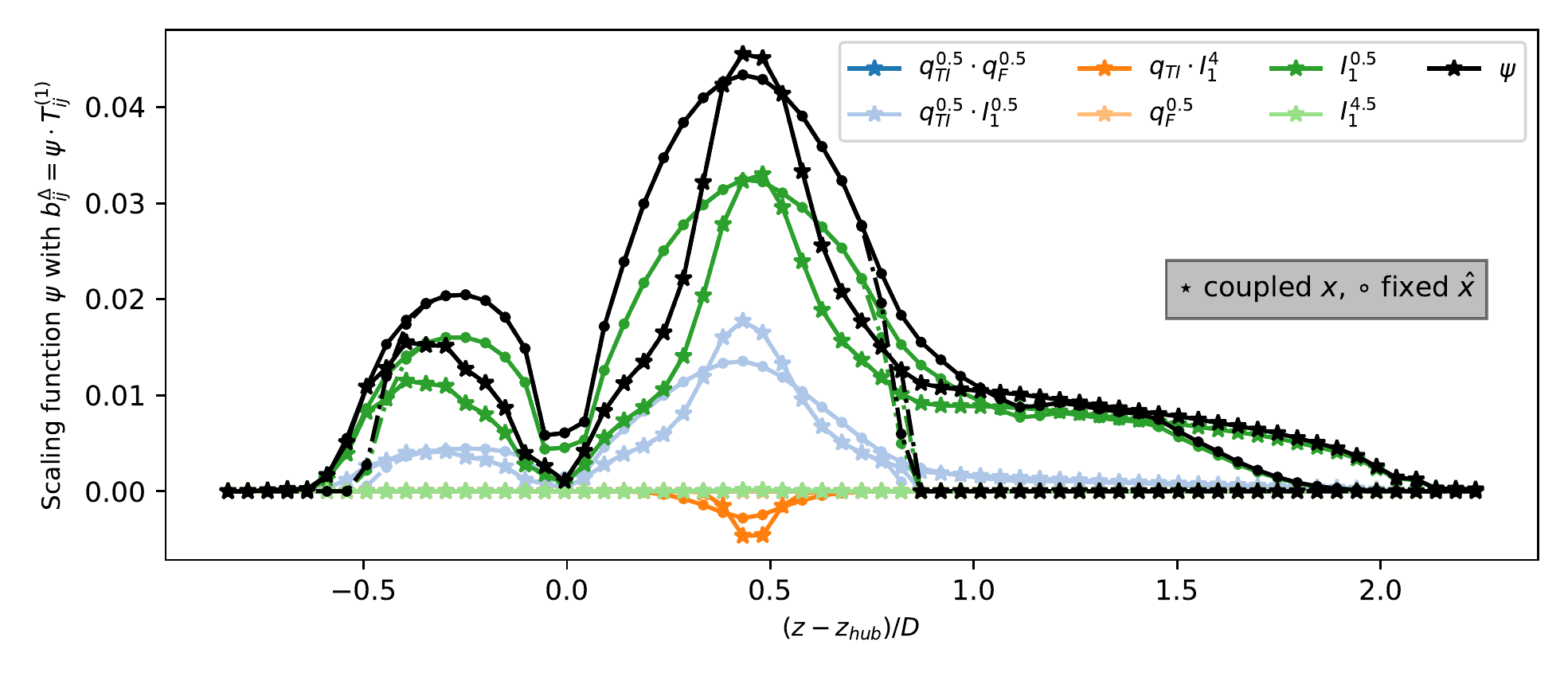} 
\caption{Visualization of (top) the features used to construct the anisotropy correction and (bottom) the the scaling function $\psi$ two diameters downstream of the second turbine of case A for the two most simple linear correction models and the one-term classifier. The $\star$ symbol means that term is calculated based on the learned and fully coupled term, the $\bullet$ symbol is the learned term calculated from the frozen invariants}
\label{fig:limBdelta}
\end{figure}

Figure \ref{fig:limBdelta} illustrates the working mechanism for the eddy viscosity limiter in a bit more detail. From the invariants in the top plot \ref{fig:limBdelta}(a) one can see that the velocity gradient based invariants tend to have sharper gradients in the fully coupled simulations. This is then also reflected in the shape of the correction term, as visible in the bottom plot \ref{fig:gradScatter}(b). All the dominant terms at that location of the flow field are based on the first invariant $I_1$ and hence also the full correction term has sharper gradients and a slightly higher maximum value. Since the anisotropy correction is further differentiated in the Navier-Stokes equations, this sometimes lead to an unstable coupling loop.

\begin{figure}[h!]
\centering
\includegraphics[width=\textwidth]{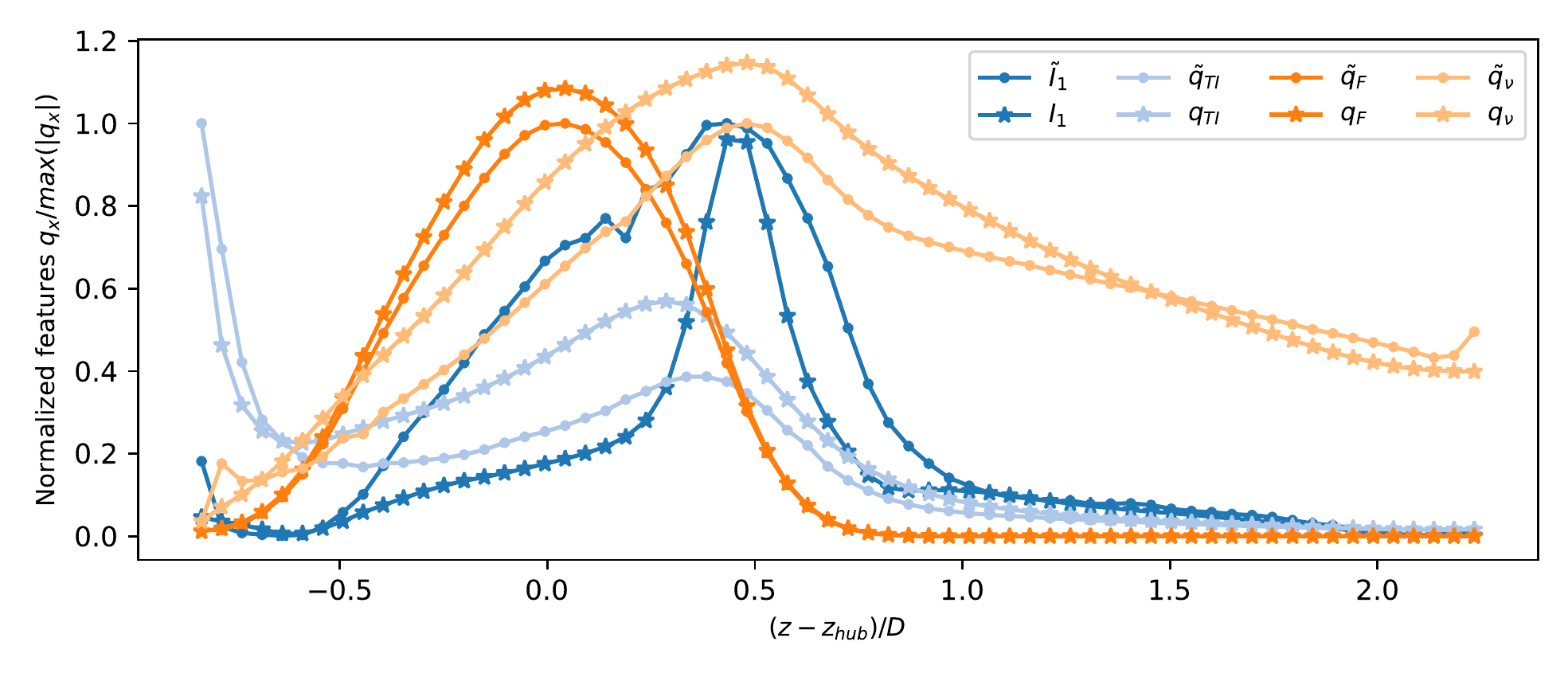} 
\includegraphics[width=\textwidth]{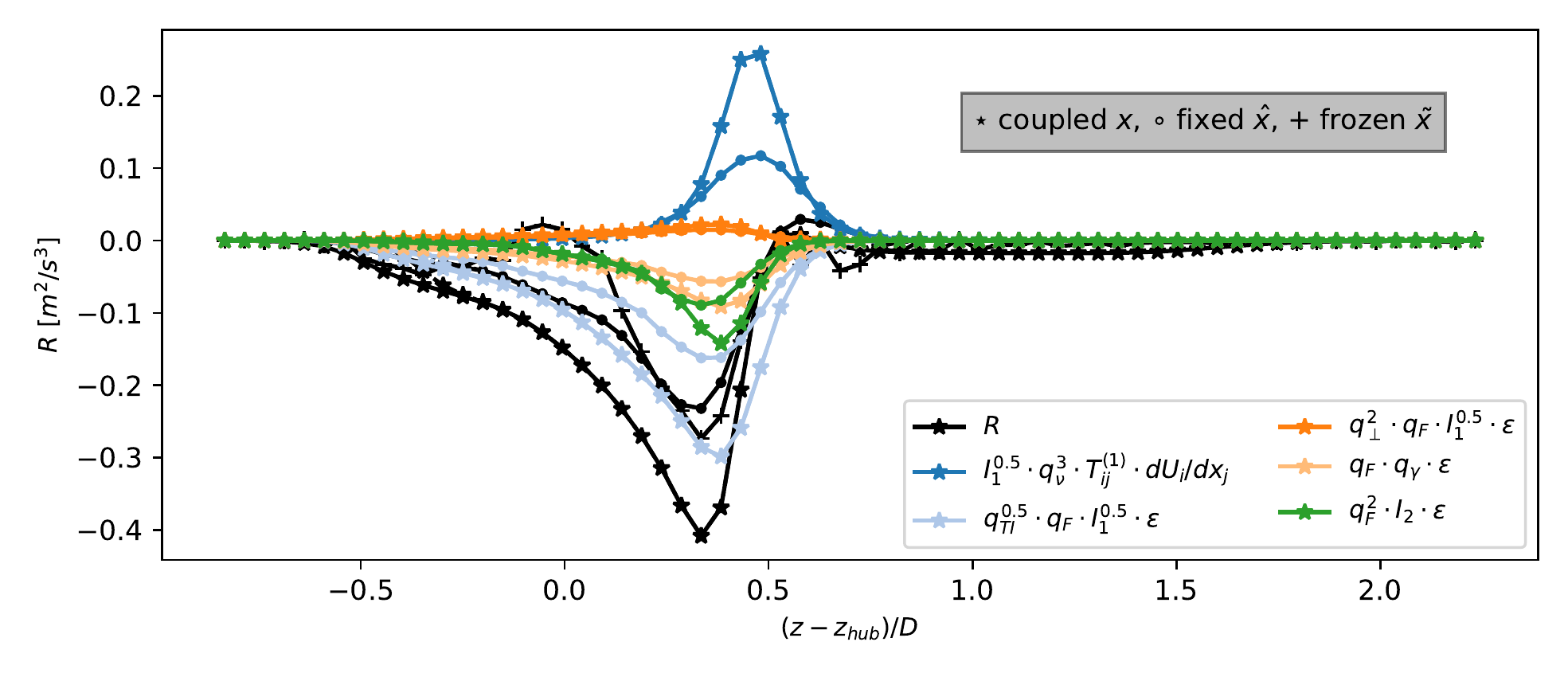} 
\caption{Visualization of (bottom) the TKE production correction and (top) the features  used to construct it at the rotor disk of the second turbine of case A for the two most simple linear correction models and the one-term classifier. The $\star$ symbol means that term is calculated based on the learned and fully coupled term, the $\bullet$ symbol is the learned term calculated from the frozen invariants so the fixed term, and the $+$ symbol is the frozen term.}
\label{fig:limR}
\end{figure}

Figure \ref{fig:limR} further highlights the second instability mechanism that was observed for the turbulent kinetic energy production correction term in the area of the flow field where the rotor forcing was non-zero. The correction term is only applied in areas where there is rotor forcing. Again the top plot \ref{fig:limR}(a) shows the different features for the frozen and the coupled simulations where one can again see that the gradient based invariants show larger gradients for the coupled simulations. The bottom plot \ref{fig:limR}(b) then shows how this affects the full correction term which is again dominated by terms related to the first invariant $I_1$. This worked well for all the simulations that were affected by this instability and usually the term is only active in a handful of cells.

Ideally such limiter functions do not need to be used in conjunction with the data-driven models, however in this case it was necessary to ensure reliable convergence of the models across the cases. Judging from these observations it may be advisable to use a logarithmic function in our library to limit the effect a growing flow gradient can have on the corrections. Likely this will decrease the accuracy of the models and could potentially make them less generic. However, it may also lead to more stable models. This should be explored further in a follow-up study.

\subsection{Flow field with learned correction terms}
\label{sec:learned}

\begin{figure}[h!]
\includegraphics[width=\textwidth]{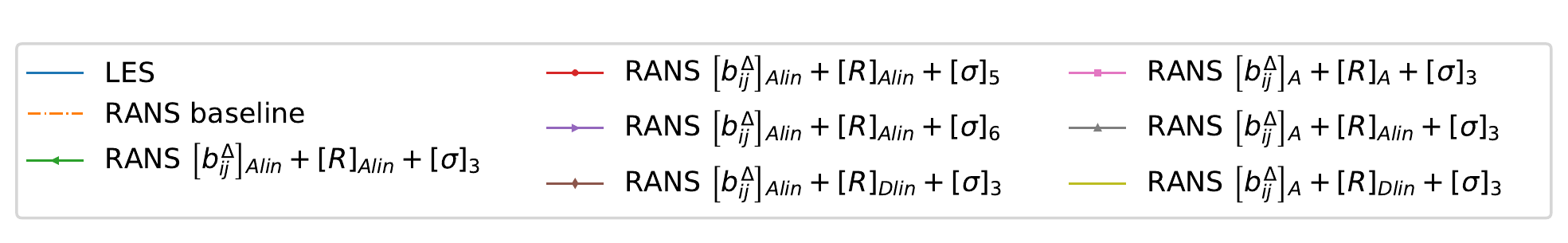} \\ 
\centering
\includegraphics[width=\textwidth]{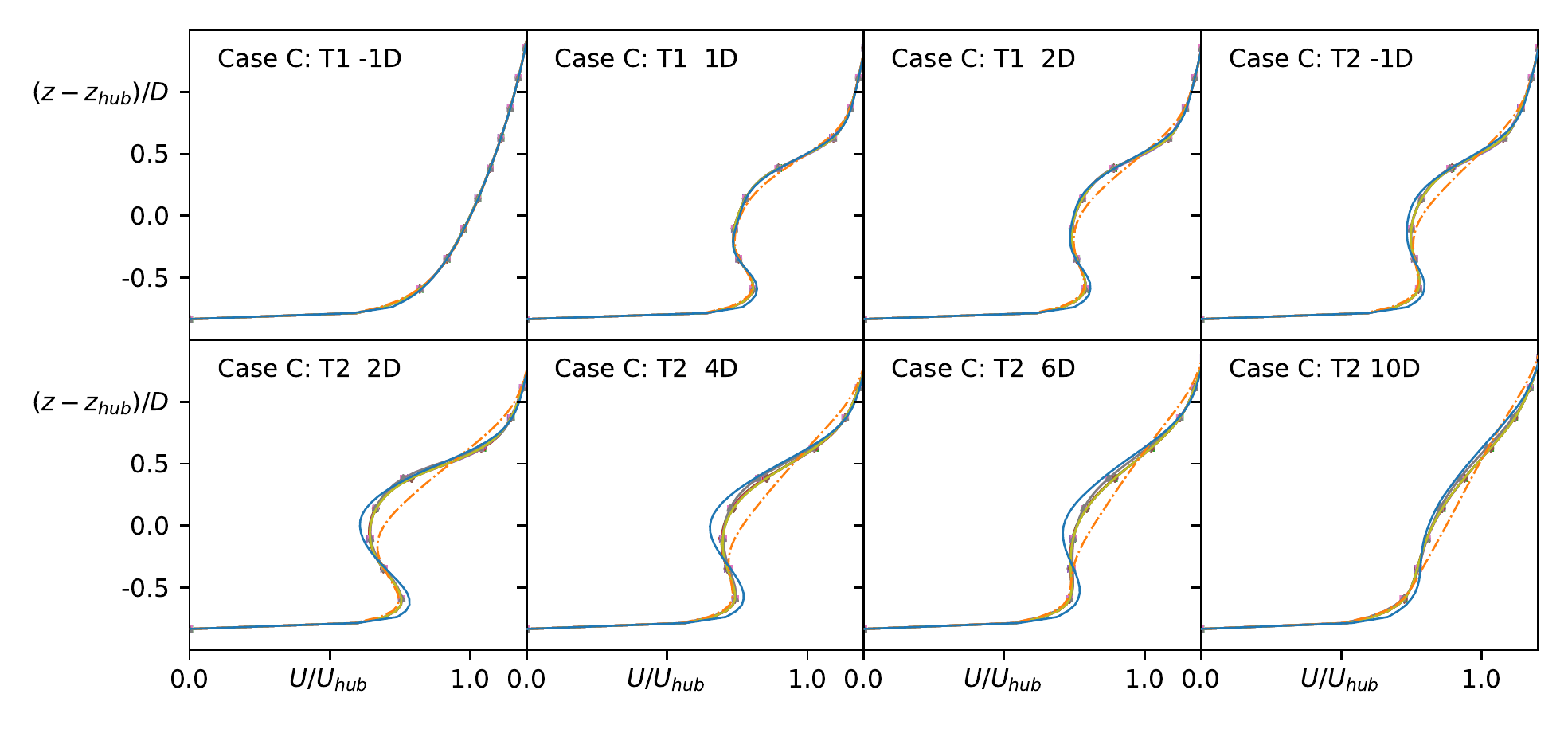} 
\includegraphics[width=\textwidth]{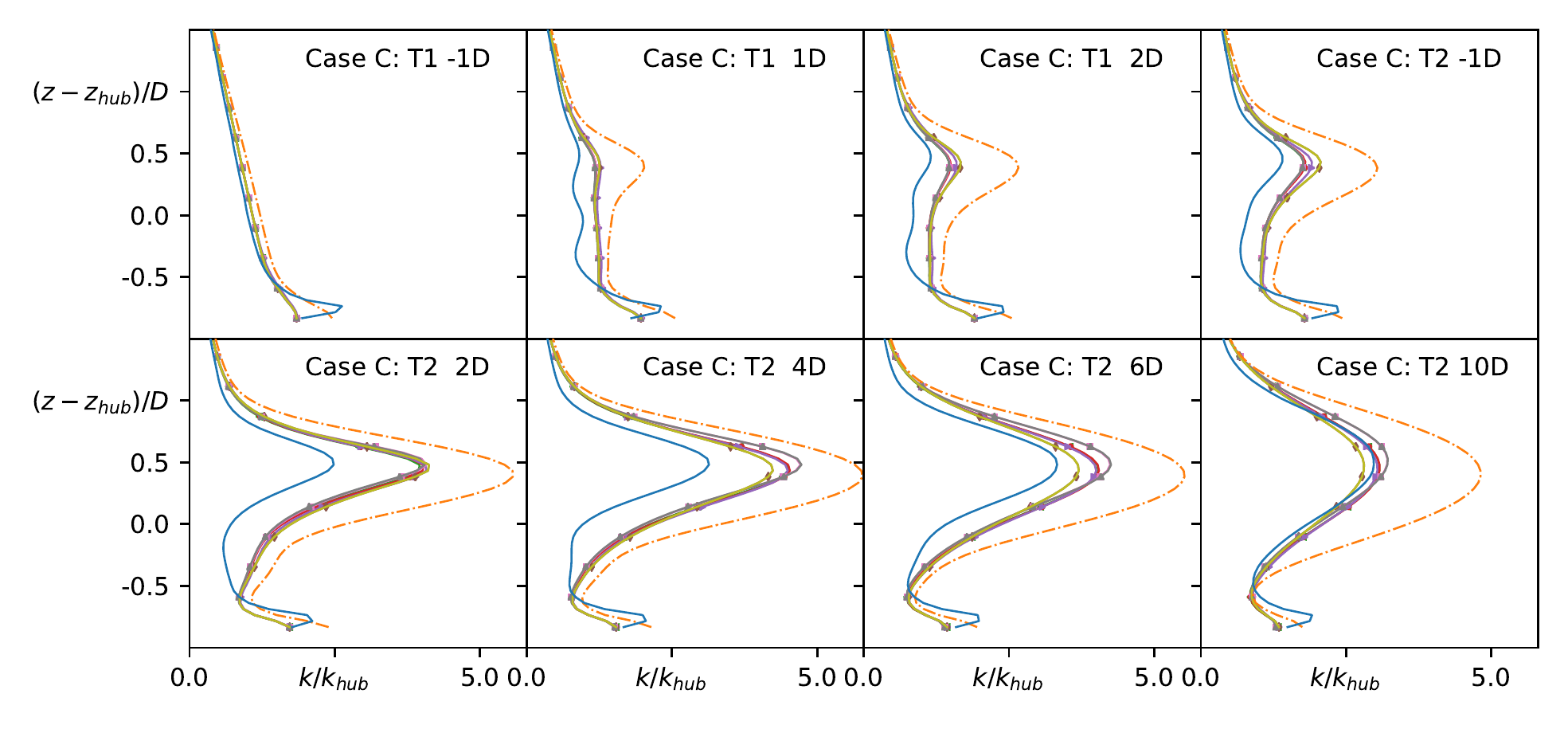} 
\caption{\label{fig:caseCcoupledUk} Comparison between LES, RANS baseline and corrected RANS models via vertical slices of the velocity and turbulent kinetic energy field up and downstream of the rotor plane for the top two turbines of case C.}
\end{figure}

Finally, all the pieces are put together. A limited selection of correction terms and classifiers are coupled with the flow solver in the fully-coupled configuration $(b_{ij}^\Delta, R, \sigma)$. Figure \ref{fig:caseCcoupledUk} shows the associated results for the test data-set C. Additional figures are presented in appendix. Only case C is shown here as the results are generally similar for all the cases.

The model selection is briefly motivated below:
\begin{itemize}
    \item Classifier $\sigma$: The most simple $\left[\sigma\right]_5$ and two intermediately complex $\left[\sigma\right]_3$, $\left[\sigma\right]_6$ classifiers were tested to see if they can yield some improvement over the most simple classifier. The more complex classifiers did not yield an improvement in the partially coupled runs and hence were not further investigated here.
    \item Anisotropy correction $b_{ij}^\Delta$: The simplest linear $\left[b_{ij}^\Delta\right]_{Alim}$ and nonlinear  $\left[b_{ij}^\Delta\right]_{A}$ model was picked, because the spread between the different models was very small in the partially coupled runs.
    \item TKE production correction $R$: The simplest $\left[R\right]_{Alim}$ and most complex linear $\left[R\right]_{Dlim}$ model were used. The same was also tried for the nonlinear models, but the terms were not robust. Hence, since the difference between linear and nonlinear terms was small in the partially coupled simulations, this was considered sufficient.
\end{itemize}

When looking at the results for the different cases, the spread as compared to the runs where the models were coupled separately did not increase. This indicates that there is no strong interaction between the two correction terms which is reassuring. Further, all the correction models yielded a significant improvement over the baseline model for both training and test data-sets. As already seen previously, there is basically no spread between the correction models for the first turbine, but the spread increases for the downstream turbines. This could be due to either the model formulation itself or the error accumulation as the flow field progresses further downstream.

Additionally, in these results, the eddy viscosity limiter was active on average in about 1500 cells mainly in the upper part of the near wake, and the form error limiter was active on average in about 4000 cells mainly in the center of the rotor disk. Given that the total number of cells in the domain is around 3 million, the limiter is seldom used, which is deemed acceptable.

Finally, in addition to figure \ref{fig:modelTerms} where there are tables of the different model terms, the most simple model formulations are written out in full for illustration:

\begin{subequations}
\begin{equation}
\begin{split}
    \left[ b_{ij}^\Delta \right]_{Alin} = \ \
    [  & 1.62 \cdot 10^{-1} \cdot q_{TI}^{1/2} \cdot q_F^{1/2} \\
	+ & 4.84 \cdot 10^{-3} \cdot q_{TI}^{1/2} \cdot I_1^{1/2} \\
	- & 1.90 \cdot 10^{-11} \cdot q_{TI} \cdot I_1^4 \\
	+ & 2.51 \cdot 10^{-2} \cdot q_F^{1/2} \\
	+ & 2.00 \cdot 10^{-3} \cdot I_1^{1/2} \\
	+ & 1.49 \cdot 10^{-15} \cdot I_1^{9/2} ] \cdot T_{ij}^{(1)}
\end{split}
\end{equation}

\begin{equation}
\begin{split}
\left[ R \right]_{Alin} =  \ \
    &
     8.06 \cdot 10^{-5} \cdot I_1^{1/2} \cdot q_\nu^3 \cdot
    k  \cdot T_{ij}^{(1)} \frac{\partial u_i}{\partial x_j} + \\
  [
    -& 2.91 \cdot 10^{1} \cdot q_{TI}^{1/2} \cdot q_F \cdot I_1^{1/2} \\
  + & 4.28 \cdot 10^{-1} \cdot q_\perp^2 \cdot q_F \cdot I_1^{1/2} \\
  - & 1.22 \cdot q_F \cdot q_\gamma \\
  + & 2.30 \cdot q_F^{2} \cdot I_2
  ] \cdot \epsilon
\end{split}
\end{equation}

\begin{equation}
    \left[ \sigma \right]_{5} = 1/\left(1+\exp{\left(
    - 205.041112 \cdot q_{TI} \cdot q_\gamma^{1/2} \cdot q_\nu^{1/2}
    + 9.01862802
    \right)}\right)
\end{equation}
\end{subequations}

The magnitude of the different components of the terms can be misleading, because the range of magnitude of the different terms is quite large. Figure \ref{fig:limR} shows for example that, although the first term of the $R$ correction has a small component, it is one of the second largest term at that specific location of the domain.

\FloatBarrier

\subsection{Comparison with models without classifier}\label{sec:wwoCL}

\begin{figure}[h!]
\centering
\includegraphics[width=\textwidth]{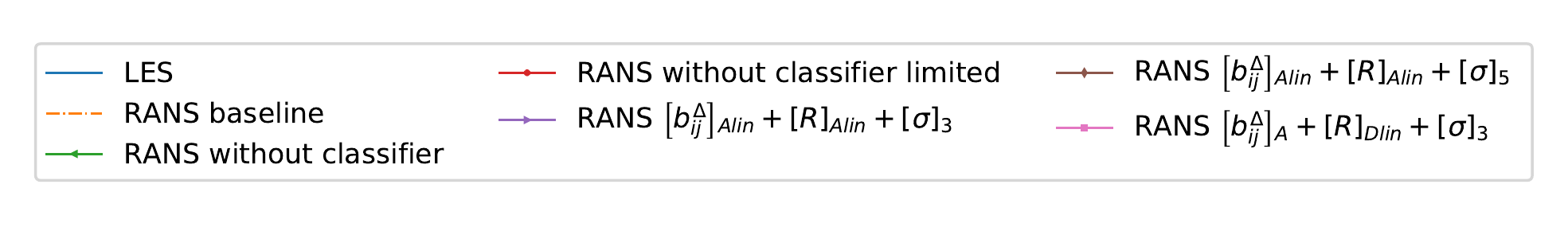} \\ 
\includegraphics[width=\textwidth]{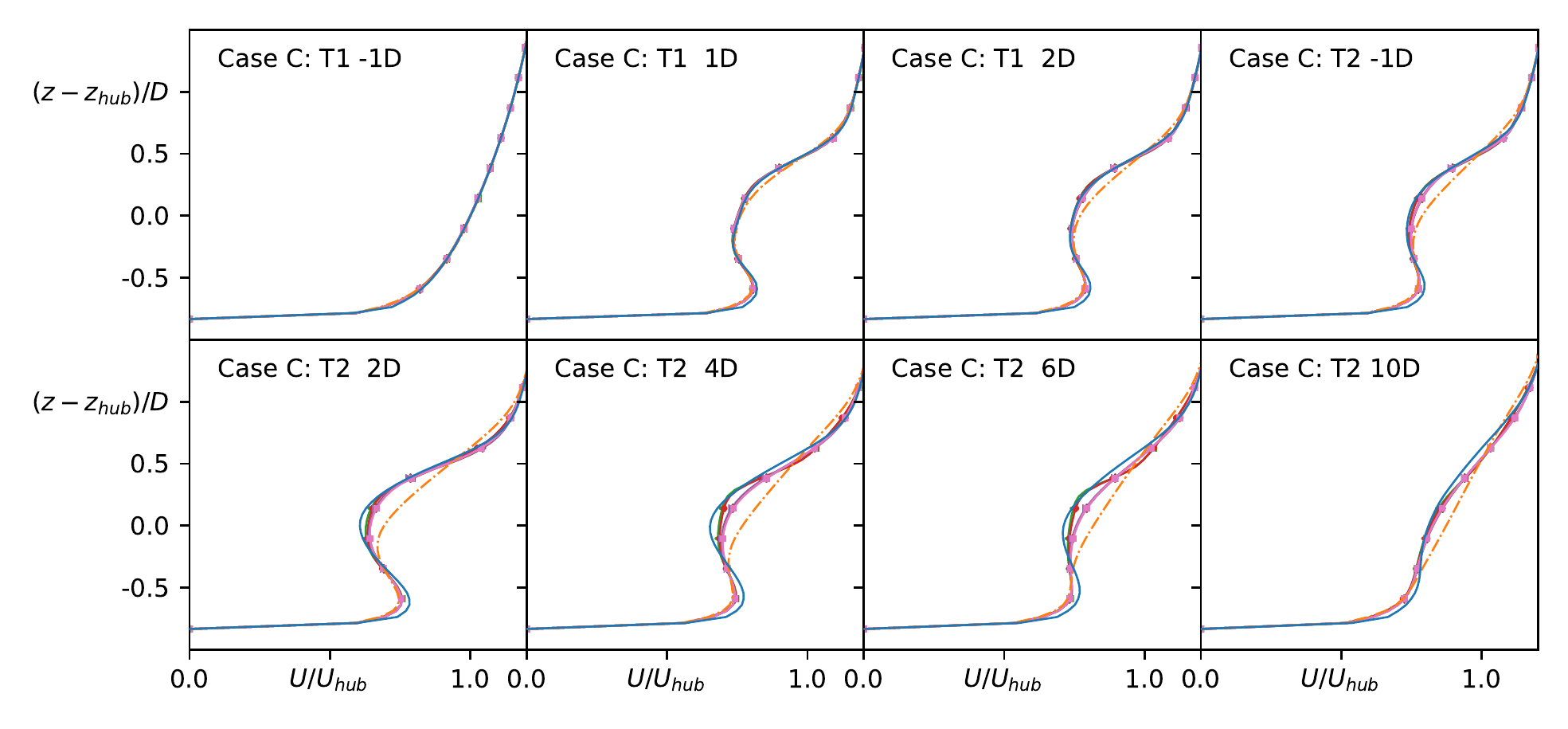} 
\includegraphics[width=\textwidth]{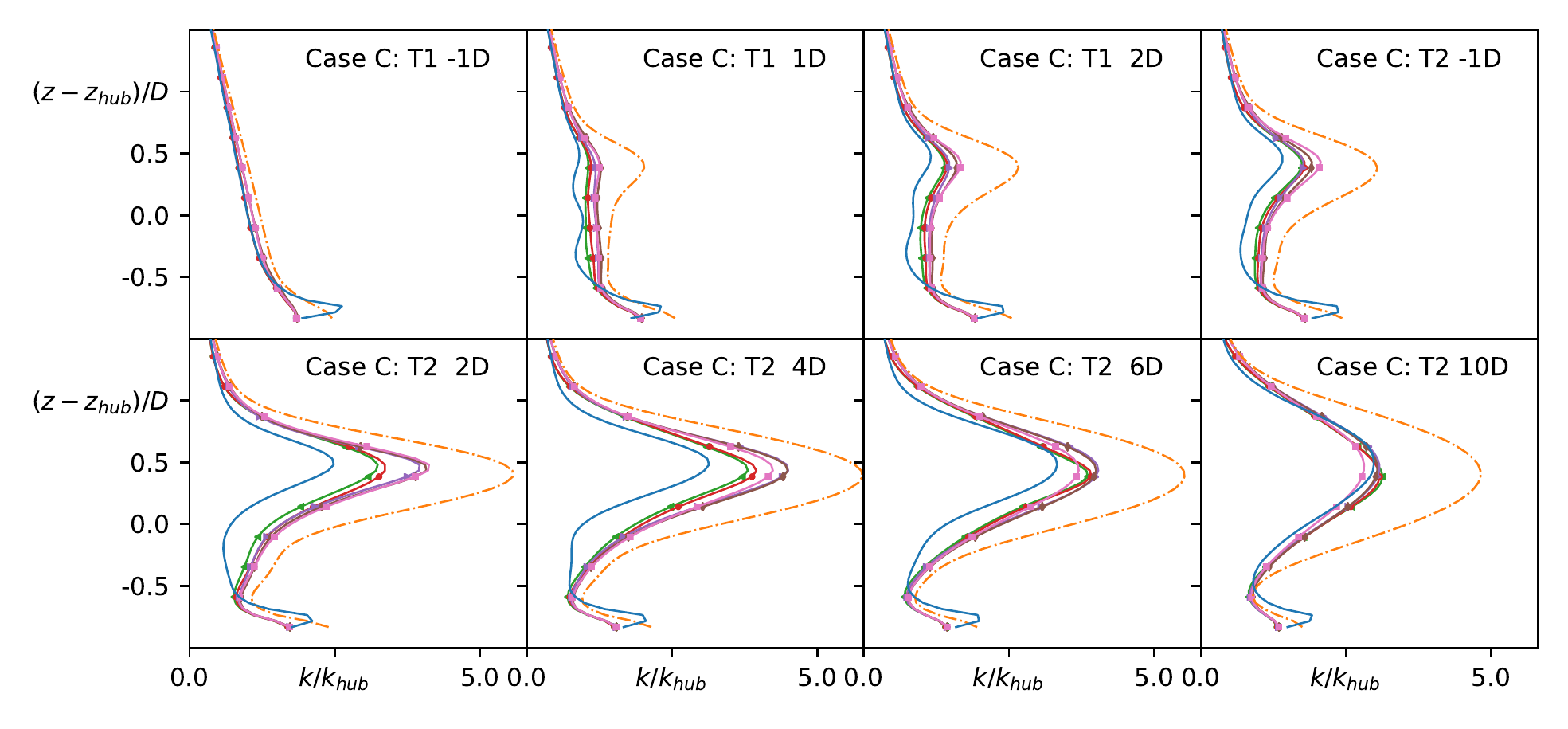} 
\caption{\label{fig:caseCcoupledUkComp} Comparison between LES, RANS baseline and corrected RANS models with and without the classifier via vertical slices of the velocity and turbulent kinetic energy field up and downstream of the rotor plane for the top two turbines of case C.}
\end{figure}

To wrap up, Figure \ref{fig:caseCcoupledUkComp} presents a comparison between models with and without the classifiers for case C. Figures \ref{fig:caseAcoupledUkComp} and \ref{fig:caseBcoupledUkComp}, respectively, show the results for case A and B and can be found in the appendix in f. For the sake of comparison, the limiters were also applied to the models without classifier and they resulted in small differences in the TKE profiles.

There are some differences between the models with and without classifier. They mainly manifest themselves in the wake of the downstream turbines and are more pronounced for the turbulent kinetic energy than for the velocity field. However, there is no clear tendency and the differences depend on the test case.

Additionally, it seems that the differences between the models with and without classifiers affect the training procedure and the numeric stability, but not necessarily the final results. Generally the training of the models was faster with the classifier, because the size of the dataset could be drastically reduced. However, as already discussed before, the models with the classifier were also more prone to numerical instability making them less robust.

Another decisive advantage of using a classifier is that simpler correction models can be used. Generally, there is some overlap between the most important terms for both corrections, but the corrections obtained with the classifier are simpler.

\FloatBarrier

\section{Conclusions}\label{sec:conclusions}
Previously, the authors demonstrated the ability of the k-corrective frozen RANS approach to yield significant improvement over the baseline $k$-$\epsilon$ model on the wind tunnel scale dataset with multiple wind turbines \cite{steiner2020a,steiner2020b}. Now the authors have expanded the approach through addition of a classifier that makes sure that the corrections are only applied selectively.

The results show that the use of a classifier results in almost the same amount of improvement over the baseline model as for the models derived without the use of a classifier. However, it also offers some advantages: (i) the resulting models are simpler and (ii) the computational effort for the training of the correction term is reduced by a factor of about 6 for this particular data-set. Also, since the training data-set becomes smaller with the classifier, trying out different learning strategies or feature sets becomes more accessible.

The study also shows that the addition of the classifier has some drawbacks: (i) it makes the models less numerically stable and (ii) it requires an additional training step for the classifier which has to be solved using logistic regression. The instability issues were resolved through implementation of limiters on the eddy viscosity and on the rotor forcing in the areas where the rotor forcing is active. It was shown that these limiters were effective despite being active in only a handful of cells. A more detailed analysis of the instability mechanisms was traced back to a positive coupling loop between the gradient based terms in the corrections. For future research, the authors propose that this can be remedied by limiting the exponents of the gradient based correction terms, by using more features that are based on non-differentiated values, and by using gradient limited formulations of the features such as the logarithmic function. Possibly, this coupling loop could also be a characteristic of the data-set itself, since both correction terms remove energy almost everywhere in the flow field. It should also be noted that the same positive coupling effects were observed in the models without a classifier, but for the resolution used in this study they did not lead to diverging simulations.

Overall, the addition of a classifier to the data-driven turbulence modeling approach SpaRTA has proven beneficial. There are some drawbacks with respect to robustness of the models, but they pertain to shortcomings of NLVEM in general, and are hence worth further investigation in a follow-up study.

\begin{appendices}
\section{Input features}\label{app:Features}

\begin{table}[!htbp]
    \caption{Physics interpreted flow features. For each feature $q_i$ the physical description is denoted including the raw feature with its normalization. The features that are not Galilean invariant are marked with ${\dagger}$.}
	\begin{tabularx}{0.95\linewidth}{x b s s}
    \toprule \midrule
        ID & Description & Raw feature & Normalization\\ \midrule
  $q_Q$ & Ratio of excess rotation rate to strain rate (Q criterion)
    & $\frac{1}{2}(\left\|\boldsymbol{\Omega}\right\|^2 - \left\|\textbf{S}\right\|^2)$
    & $\left\|\textbf{S}\right\|^2$ \\
  $q_{TI}^{{\dagger}}$ & Turbulence intensity
    & $k$
    & $\frac{1}{2}U_iU_i$ \\
  $q_{ReD}$ & Wall distance based Reynolds number
    & $\frac{\sqrt{k}d}{50\nu}$
    &   - \\
  $q_{\partial p \partial s}^{{\dagger}}$ & Pressure gradient along streamline
    & $U_k\frac{\partial P}{\partial x_k}$
    & $\sqrt{\frac{\partial P}{\partial x_j}\frac{\partial P}{\partial x_j}U_i U_i}$ \\
  $q_T$   &   Ratio of mean turbulent to mean strain time scale
    & $\frac{k}{\varepsilon}$
    & $\frac{1}{\left\|\textbf{S}\right\|}$ \\
  $q_{\nu}$ & Viscosity ratio
      & $\nu_t$
      & $100\nu$ \\
  $q_{\perp}^{{\dagger}}$ & Nonorthogonality between velocity and its gradient
    & $|U_iU_j \frac{\partial U_i}{\partial x_j}|$
    & $\sqrt{U_lU_lU_i\frac{\partial U_i}{\partial x_j}U_k \frac{\partial U_k}{\partial x_j}}$ \\
  $q_{\mathcal{C}_k/\mathcal{P}_k}^{{\dagger}}$ & Ratio of convection to Boussinesq production of TKE
    & $U_i\frac{dk}{dx_i}$
    & $|\overline{u_j'u_k'}S_{jk}|$ \\
  $q_{\tau}$ & Ratio of total to normal Boussinesq Reynolds stresses
    & $||\overline{u_i'u_j'}_{BS}||$
    &  $k$ \\
  $q_{\gamma}$ & Shear parameter
    & $\left\|\frac{\partial U_i}{\partial x_j}\right\|$
    & $\frac{\varepsilon}{k}$ \\
  $q_{F}^{{\dagger}}$ & Actuator forcing
  	& $\left\|F_{cell}\right\|$
    & $\frac{1}{2} \rho_0 A_{cell} \left\| U \right\|^2$ \\
	\midrule \bottomrule
	\end{tabularx}
    \label{tab:physicalFeaturesAll}
\end{table}

\begin{table}[!htbp]
    \centering
    \caption{Invariant bases, number of symmetric and antisymmetric tensors for each invariant are indicated by $n_s$ and $n_A$, respectively. The invariant bases are the trace of the tensors listed. The asterisk on a invariant bases indicates that also the cyclic permutation of the antisymmetric tensors are included.}
    \def\arraystretch{1.25}
    \begin{tabular}{ccc} \toprule[1pt]\midrule[0.3pt]
        $(n_S,n_A)$ & Feature index & Invariant bases \\ \midrule
        $(1,0)$   &   1-2 &   $\textbf{S}^2$, $\textbf{S}^3$    \\
        $(0,1)$   &   3-5 &   $\boldsymbol{\Omega}^2$, $\textbf{A}_p^2$, $\textbf{A}_k^2$      \\
        $(1,1)$   &   6-14 &   $\boldsymbol{\Omega}^2 \textbf{S}$, $\boldsymbol{\Omega}^2 \textbf{S}^2$, $\boldsymbol{\Omega}^2 \textbf{S} \boldsymbol{\Omega} \textbf{S}^2$   \\
           &    &   $\textbf{A}_p^2\textbf{S}$, $\textbf{A}_p^2\textbf{S}^2$, $\textbf{A}_p^2 \textbf{S} \textbf{A}_p \textbf{S}^2$    \\
           &   & $\textbf{A}^2_k\textbf{S}$, $\textbf{A}^2_k\textbf{S}^2$ , $\textbf{A}^2_k\textbf{S}\textbf{A}_k\textbf{S}^2$ \\
        $(0,2)$   &   15-17 &   $\boldsymbol{\Omega}\textbf{A}_p$, $\textbf{A}_p\textbf{A}_k$, $\boldsymbol{\Omega}\textbf{A}_k$   \\
        $(1,2)$   &   18-41 &   $\boldsymbol{\Omega}\textbf{A}_p\textbf{S}$, $\boldsymbol{\Omega}\textbf{A}_p\textbf{S}^2$, $\boldsymbol{\Omega}^2 \textbf{A}_p\textbf{S}^*$, $\boldsymbol{\Omega}^2\textbf{A}_p\textbf{S}^{2*}$, $\boldsymbol{\Omega}^2\textbf{S}\textbf{A}_p\textbf{S}^{2*}$   \\
           &    &   $\boldsymbol{\Omega}\textbf{A}_k\textbf{S}$, $\boldsymbol{\Omega}\textbf{A}_k\textbf{S}^2$, $\boldsymbol{\Omega}^2 \textbf{A}_k\textbf{S}^*$, $\boldsymbol{\Omega}^2\textbf{A}_k\textbf{S}^{2*}$, $\boldsymbol{\Omega}^2\textbf{S}\textbf{A}_k\textbf{S}^{2*}$    \\
           &   &    $\textbf{A}_p\textbf{A}_k\textbf{S}$, $\textbf{A}_p\textbf{A}_k\textbf{S}^2$, $\textbf{A}^2_p\textbf{A}_k\textbf{S}^*$, $\textbf{A}^2_p\textbf{A}_k\textbf{S}^{2*}$  \\
        $(0,3)$   &   42 &   $\boldsymbol{\Omega}\textbf{A}_p \textbf{A}_k$  \\
        $(1,3)$ & 43-47 & $\boldsymbol{\Omega}\textbf{A}_p\textbf{A}_k\textbf{S}$, $\boldsymbol{\Omega}\textbf{A}_k\textbf{A}_p\textbf{S}$, $\boldsymbol{\Omega}\textbf{A}_p\textbf{A}_k\textbf{S}^2$, $\boldsymbol{\Omega}\textbf{A}_k\textbf{A}_p\textbf{S}^2$, $\boldsymbol{\Omega}\textbf{A}_p\textbf{S}\textbf{A}_k\textbf{S}^2$ \\ \midrule[0.3pt]\bottomrule[1pt]
    \end{tabular}
    \label{tab:invariantsAll}
\end{table}

\FloatBarrier

\section{Additional figures for different test cases}

\begin{figure}[h!]
\includegraphics[width=\textwidth]{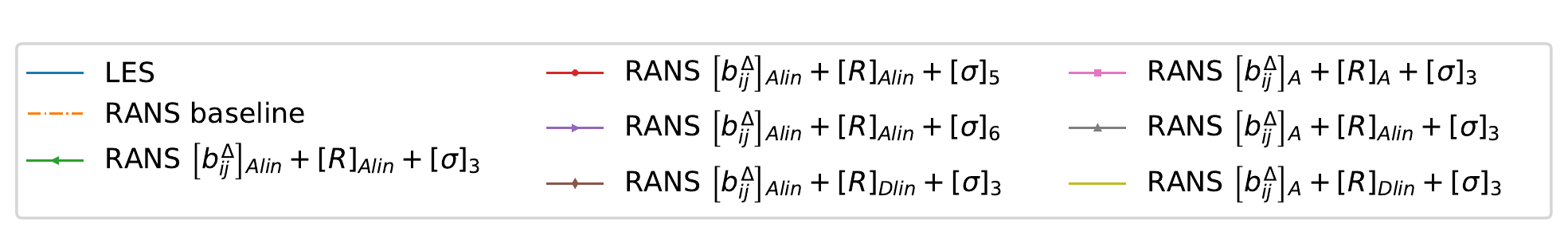} \\ 
\centering
\includegraphics[width=\textwidth]{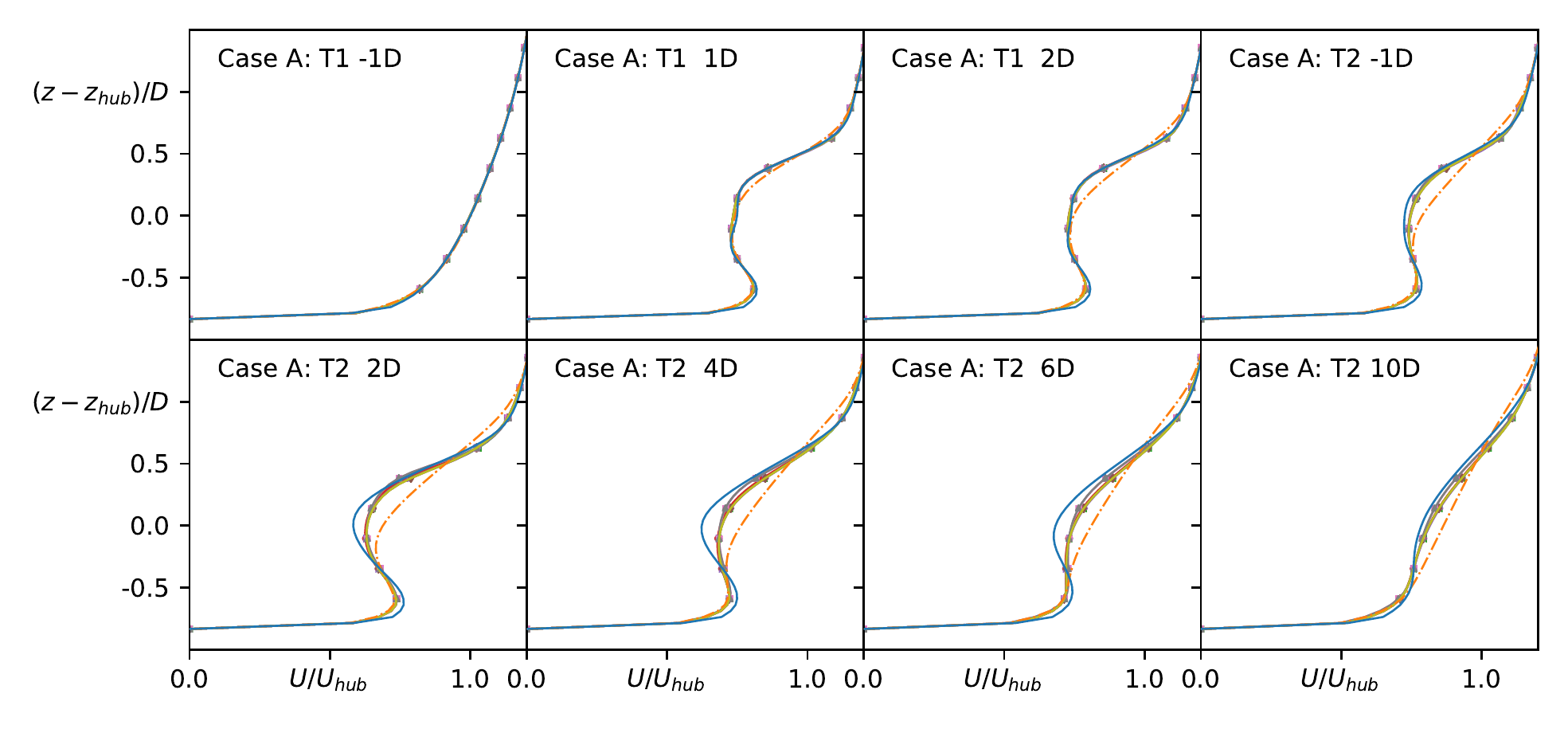} 
\includegraphics[width=\textwidth]{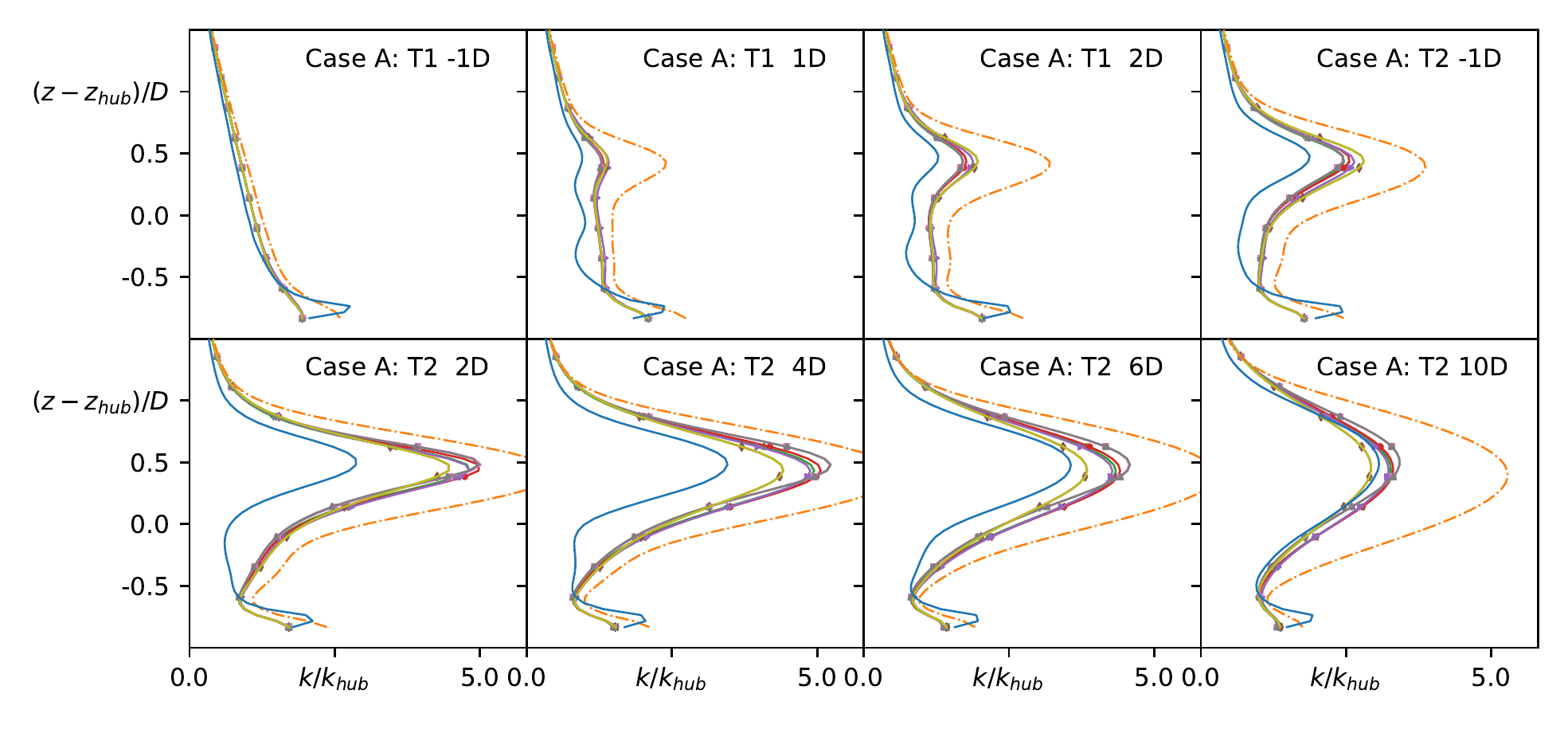} 
\caption{\label{fig:caseAcoupledUk} Comparison between LES, RANS baseline, frozen RANS and corrected RANS models via vertical slices of the velocity and turbulent kinetic energy field up and downstream of the rotor plane for the three turbines of case A.}
\end{figure}

\begin{figure}[h!]
\includegraphics[width=\textwidth]{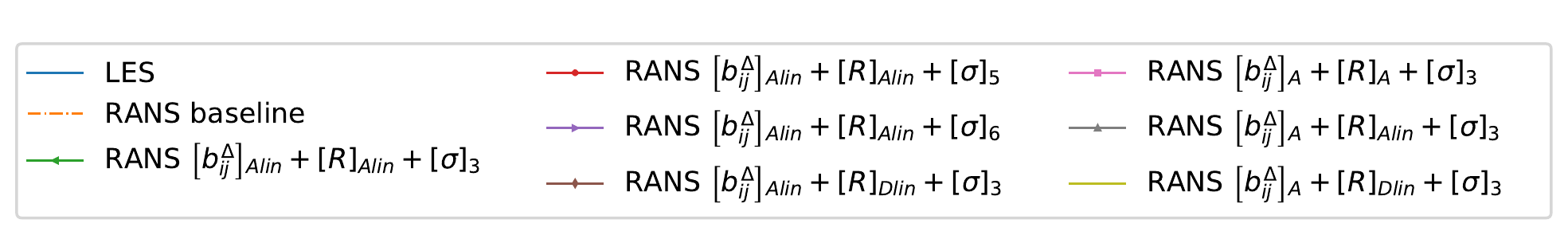} \\ 
\centering
\includegraphics[width=\textwidth]{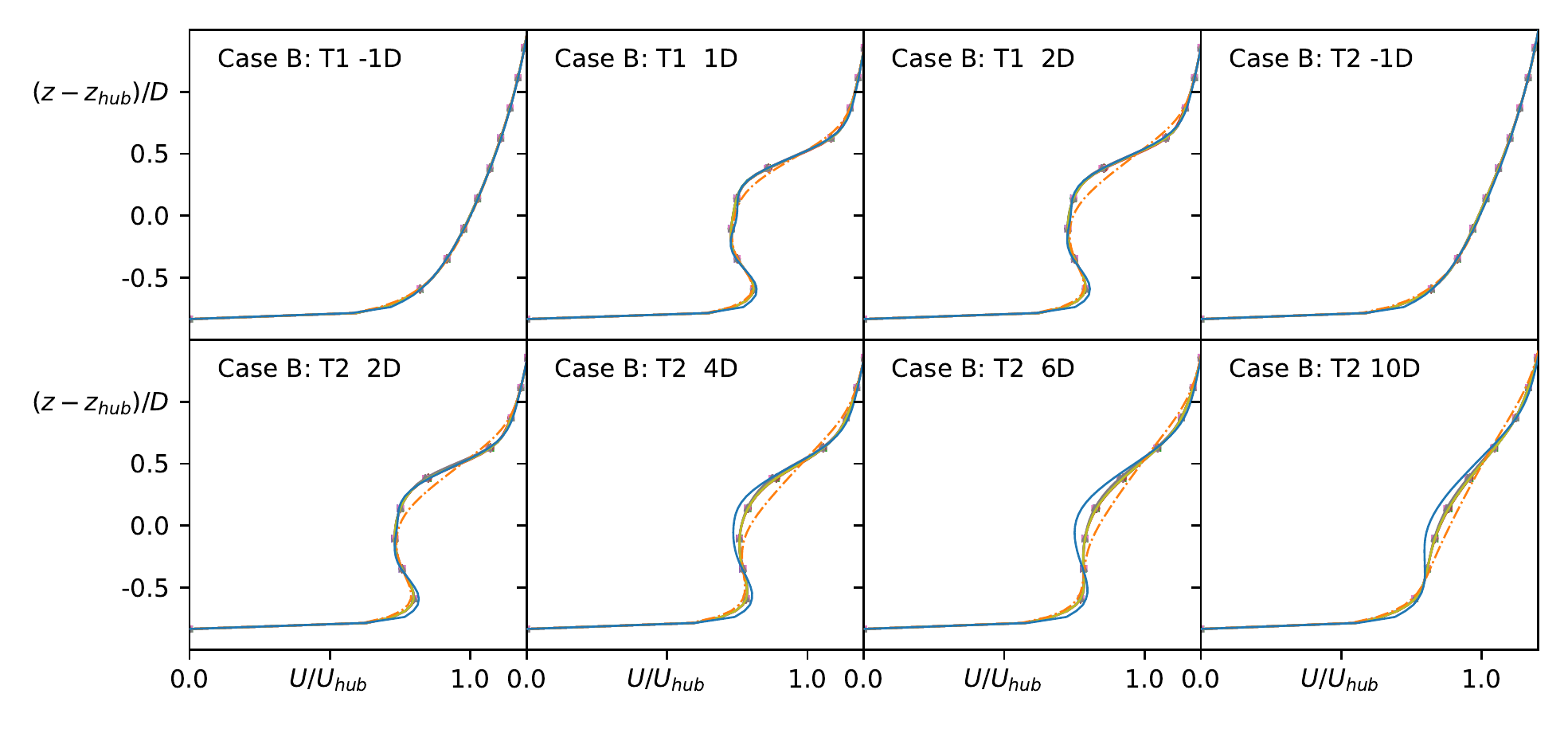} 
\includegraphics[width=\textwidth]{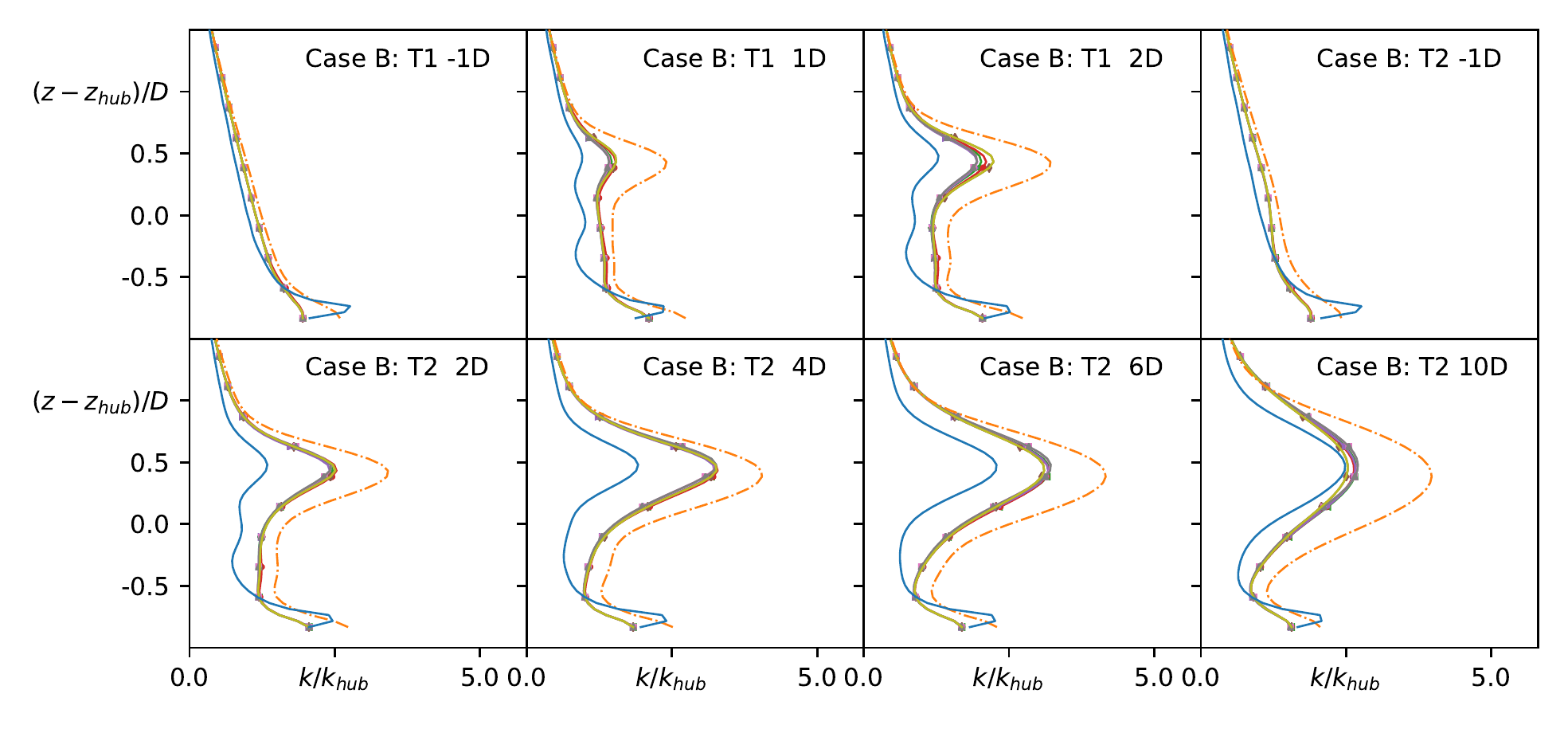} 
\caption{\label{fig:caseBcoupledUk} Comparison between LES, RANS baseline, frozen RANS and corrected RANS models via vertical slices of the velocity and turbulent kinetic energy field up and downstream of the rotor plane for the three turbines of case B.}
\end{figure}

Figures \ref{fig:caseAcoupledUk} and \ref{fig:caseBcoupledUk} show the results for the training data-set A. Figures \ref{fig:caseAcoupledUk} and \ref{fig:caseBcoupledUk} in the appendix show the results for the test data-set B.

\begin{figure}[h!]
\centering
\includegraphics[width=\textwidth]{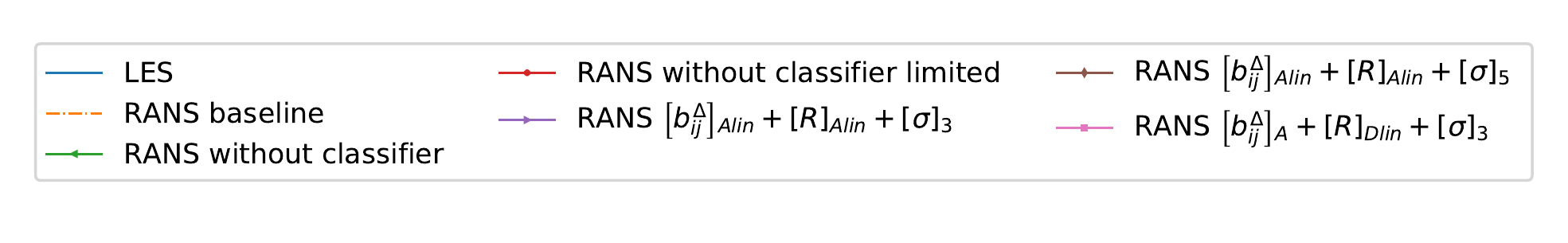} \\ 
\includegraphics[width=\textwidth]{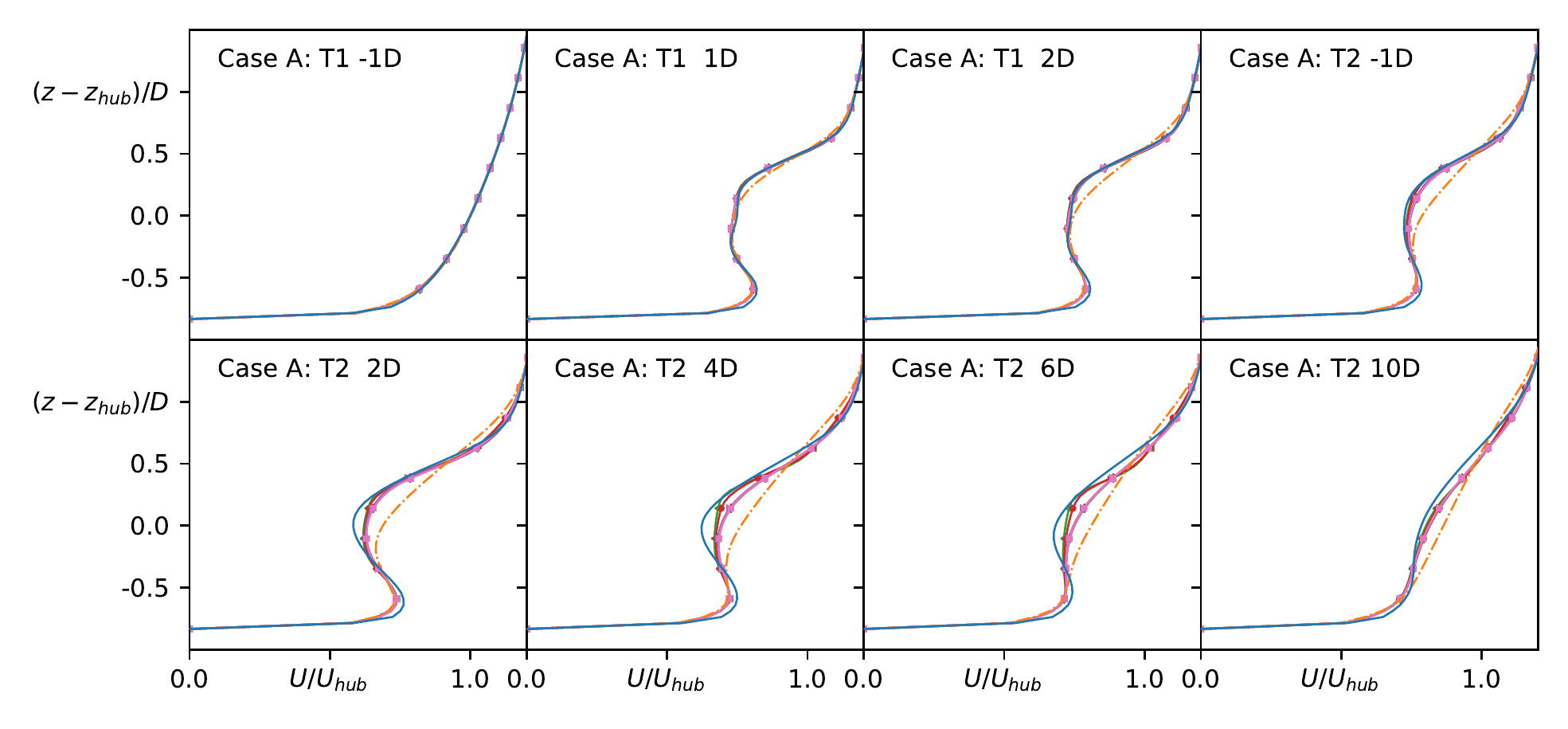} 
\includegraphics[width=\textwidth]{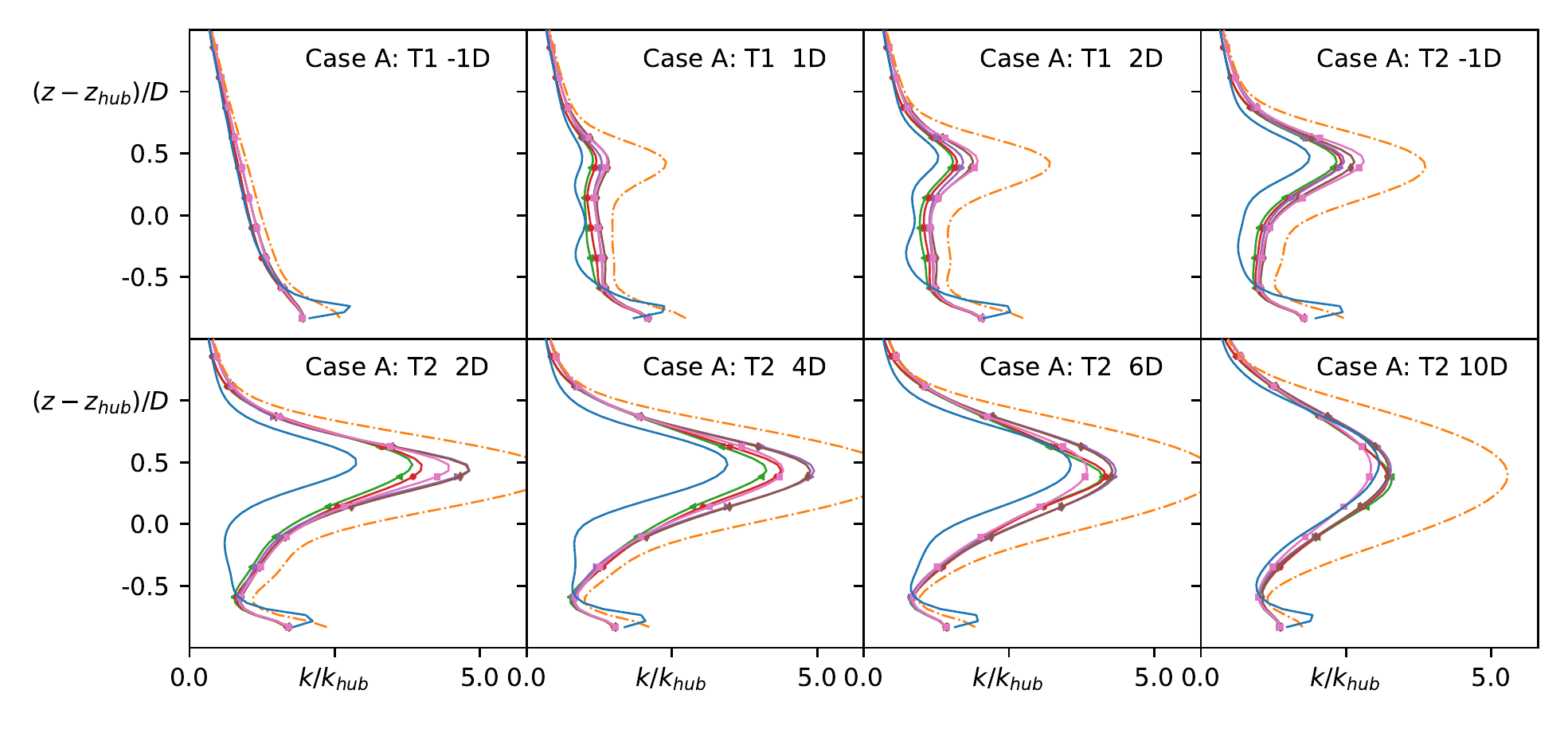} 
\caption{\label{fig:caseAcoupledUkComp} Comparison between LES, RANS baseline, frozen RANS and corrected RANS models via vertical slices of the velocity and turbulent kinetic energy field up and downstream of the rotor plane for the three turbines of case A.}
\end{figure}

\begin{figure}[h!]
\centering
\includegraphics[width=\textwidth]{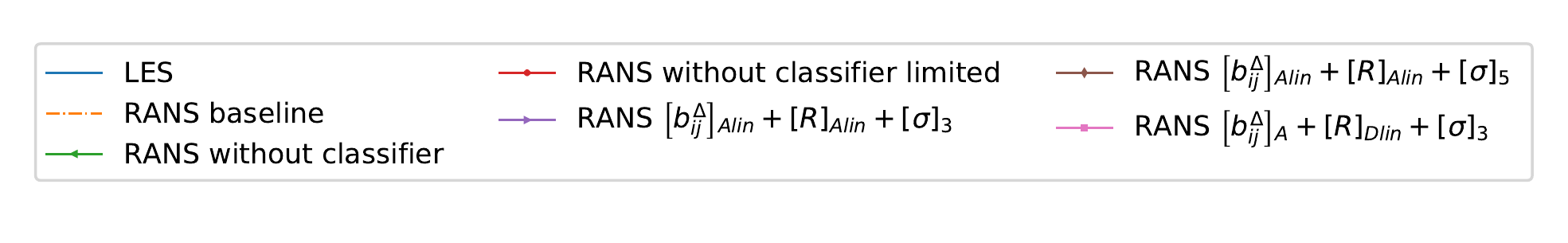} \\ 
\includegraphics[width=\textwidth]{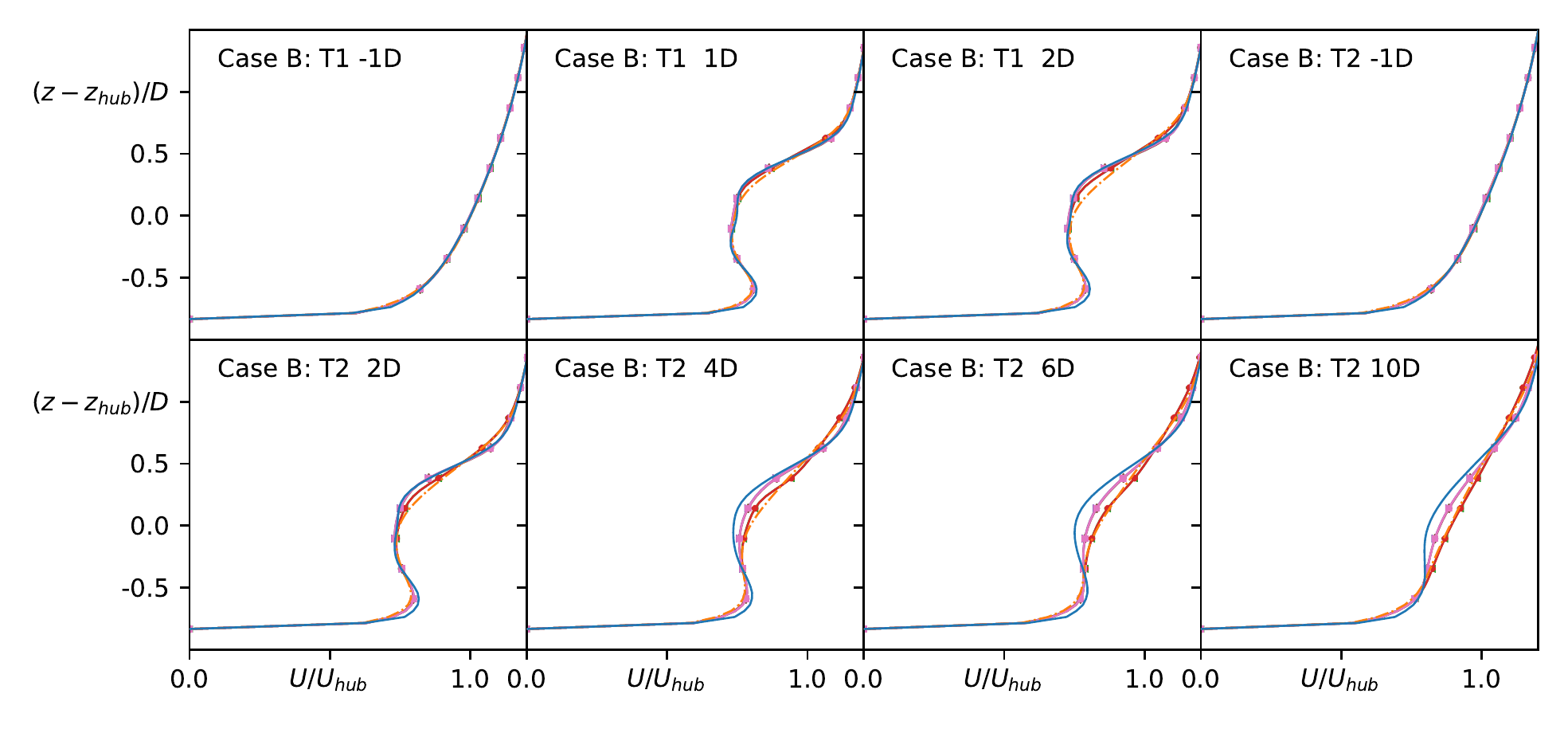} 
\includegraphics[width=\textwidth]{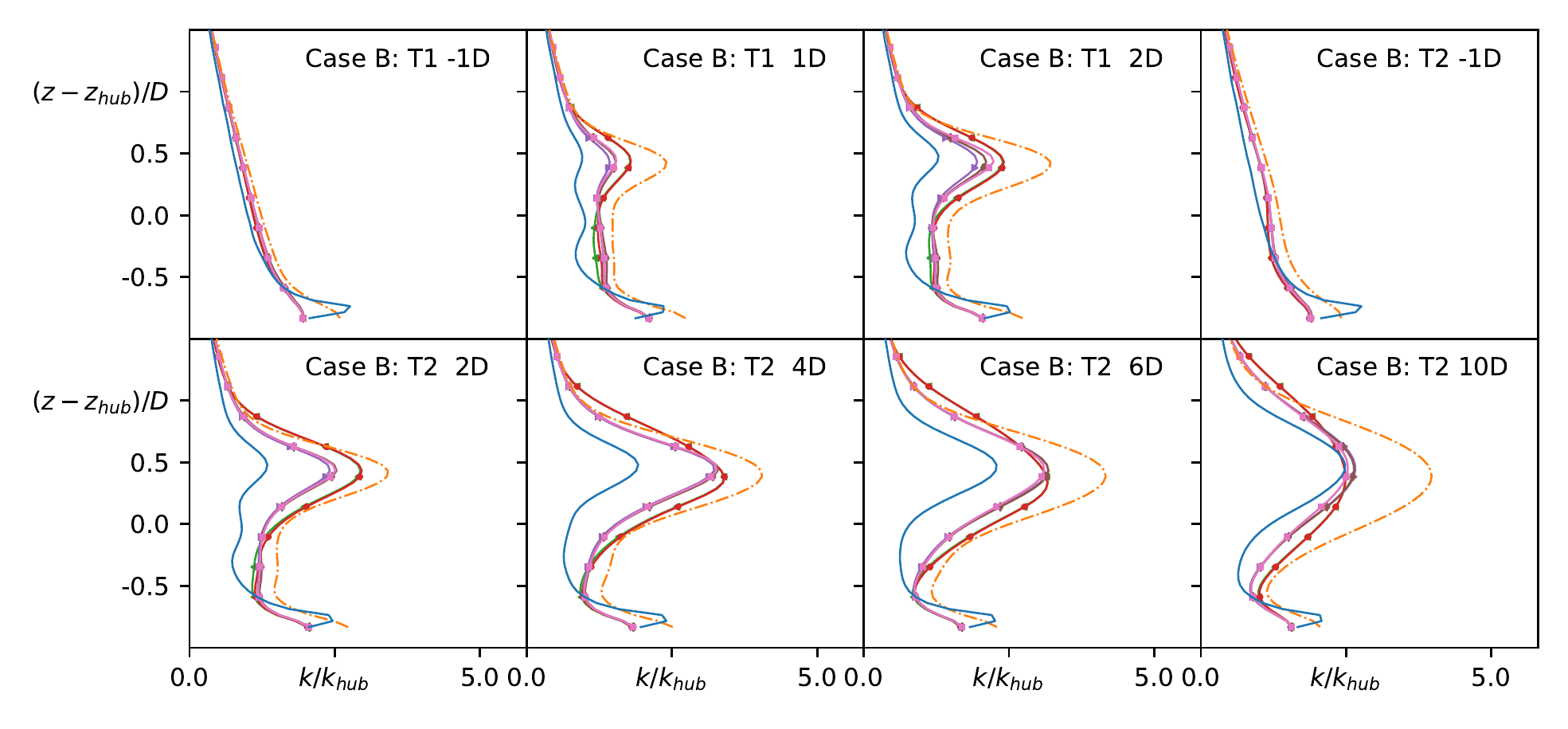} 
\caption{\label{fig:caseBcoupledUkComp} Comparison between LES, RANS baseline, frozen RANS and corrected RANS models via vertical slices of the velocity and turbulent kinetic energy field up and downstream of the rotor plane for the three turbines of case B.}
\end{figure}

\FloatBarrier

\end{appendices}

\textbf{Funding:} This study was funded by Rijksdienst voor Ondernemend Nederland (grant number TEHE116332).

\textbf{Conflict of interest:} The authors declare that they have no conflict of interest.


\end{document}